\RequirePackage{fix-cm}
\documentclass{svjour3}                     
\smartqed  
\usepackage{graphicx, subcaption, caption}
\usepackage{amsmath}
\usepackage{amssymb}
\usepackage{cancel}
\begin{document}

\title{Estimating the integral length scale on turbulent flows from the zero crossings of the longitudinal velocity fluctuation
}

\titlerunning{Estimating $\mathcal{L}$ from the variance of zero crossings}        

\author{D.O. Mora         \and
        M. Obligado 
}


\institute{D.O. Mora. \at Universit\'{e} Grenoble Alpes, CNRS, Grenoble-INP, LEGI, F-38000, Grenoble, France  and University of Washington, Department of Mechanical Engineering, Seattle, USA
           \and
           M. Obligado \at
		Universit\'{e} Grenoble Alpes, CNRS, Grenoble-INP, LEGI, F-38000, Grenoble, France
		\email{Martin.Obligado@univ-grenoble-alpes.fr}   
}

\date{Received: date / Accepted: date}

\maketitle

\begin{abstract}
	
	The integral length scale ($\mathcal{L}$) is considered to be characteristic of the largest motions of a turbulent flow, and as such, it is an input parameter in modern and classical approaches of turbulence theory and numerical simulations. Its experimental estimation, however, could be difficult in certain conditions, for instance, when the experimental calibration required to measure $\mathcal{L}$ is hard to achieve (hot-wire anemometry on large scale wind-tunnels, and field measurements), or in `standard' facilities using active grids due to the behaviour of their velocity autocorrelation function $\rho(r)$, which does not in general cross zero. In this work, we provide two alternative methods to estimate $\mathcal{L}$ using the variance of the distance between successive zero crossings of the streamwise velocity fluctuations, thereby reducing the uncertainty of estimating $\mathcal{L}$ under similar experimental conditions. These methods are applicable to variety of situations such as active grids flows, field measurements, and large scale wind tunnels.
\end{abstract}

\section{Introduction}
The integral length scale ($\mathcal{L}$) is widely interpreted as the characteristic length scale of the energy containing eddies in a turbulent flow. $\mathcal{L}$ is defined as the integral of the normalised velocity autocorrelation function; $\mathcal{L}=\int_0^\infty\rho(r)dr$, where  $\rho(r)=\langle u^\prime(x)u^\prime(x+r)\rangle/ \sigma_u^2$ \cite{tennekes1972first}, and $\sigma_u$ is the standard deviation of the streamwise velocity fluctuations $u^\prime$. Moreover, $\mathcal{L}$ is central to different attempts aiming to understand turbulence evolution and its cascading process \cite{pope_2000,vassilicos2015dissipation}. For instance, for turbulence close to a statistically homogeneous and isotropic state (HIT), the dissipation constant $C_\varepsilon=\varepsilon\mathcal{L}/\sigma_u^{3}$, depends on two large scale quantities, $\mathcal{L}$ and $\sigma_u$.

In experiments, noise and non-stationary experimental conditions can pollute the large separation values, which are denoted by increments of $r$. This prevents the computation of the integral $\int_0^\infty\rho(r)dr$ up to infinity. Therefore, the value of $\mathcal{L}$ is usually estimated by different methods. These include: integrating up to the first zero crossing \cite{o2004autocorrelation}; integrating up to a minimum value of the autocorrelation function \cite{o2004autocorrelation,tritton2012physical}; integrating up to the value where the autocorrelation falls below $e^{-1}$ \cite{o2004autocorrelation,bewley2012integral} or via standard Kolmogorov scalings. Valente \& Vassilicos \cite{valente2011decay} also propose to integrate the autocorrelation function up to a lengthscale which is about ten times $\mathcal{L}$. Moreover, Krogstad \& Davidson \cite{krogstad2011freely} suggest to apply a high-pass filter to the time signal at 0.1 Hz to counteract the effect of non-stationary low frequencies on the estimation of $\mathcal{L}$.  These previous estimations are usually accompanied by the assumption that Taylor's hypothesis holds, $r= U\tau$, where $\tau$ refers to time and $U$ is the local convective velocity.

Despite their widespread use, these approaches to estimate $\mathcal{L}$ may fail or result in ambiguities. For instance, some experimental studies using facilities that generate turbulence by means of active grids \cite{mydlarski2017turbulent} have reported that $\rho(r)$ sometimes does not decay exponentially nor cross zero \cite{puga2017normalized,MoraPRF2019}. These observations pose the problem of how to compute $\mathcal{L}$ under such conditions. Likewise, in large scale experiments \cite{gagne2004reynolds}, or  in  field  measurements,  conducting  the  equipment  calibration  procedure  could be very cumbersome, and therefore, such uncertainty would contaminate the reported values of the turbulent quantities.

To cater for the autocorrelation behaviour, Puga \& LaRue \cite{puga2017normalized} have recommended estimating the integral length scale as $\mathcal{L}=\int_0^{r_0}\rho(r)dr$ with  $r_0=U\tau_0=U\tau(\rho(\tau)=\delta)$. The parameter $\delta$ quantifies the dispersion on the estimation of $\rho(r)$. It is usually found by averaging different segments extracted from the velocity time signal. Therefore, when $\int_0^{r_0}\rho(r)dr$ is estimated by this method, $\delta$ plays an important role in the value of $\mathcal{L}$ obtained. Nevertheless, the choice of $\delta$ is ambiguous as it strongly depends on the averaging chosen for the computation of $\rho(r)$. This is not a minor issue considering the influence $\mathcal{L}$ exerts on the normalised dissipation rate constant $C_\varepsilon$, e.g.,
Mora et al. \cite{MoraPRF2019} reported $C_\varepsilon\approx 0.3$ in disagreement (by a factor of 2) with the value $C_\varepsilon\approx 0.6$ reported by Puga \& LaRue \cite{puga2017normalized} for similar values of $Re_\lambda$, despite the high degree of turbulence isotropy and turbulence homogeneity present in both experiments.

Puga \& LaRue \cite{puga2017normalized} anticipated that their method could not be general to all active grid generated flows, as it has been reported that the active grid protocol could affect the largest scales of the flow \cite{hearst2015effect,griffin2019control}. Then, the choice of $\delta$, which under this method may change between different experimental conditions or data analyses, could impact $\mathcal{L}$ and $C_\varepsilon$ making it difficult to compare different results available in the literature.

To address this problem, we study the zero crossings of $u^\prime$ for different datasets. Zero crossing analysis has been used in the past to characterise the small scales features of the flow via the Taylor microscale ($\lambda$) \cite{sreenivasan1983zero,mazellier2008turbulence,MoraPRF2019}. Given that the zero crossings of a velocity signal usually do not depend on the equipment calibration (as far as the mean velocity is known), this analysis is suitable even under challenging experimental conditions.

The first approach proposed  is solely based on the work of McFadden \cite{mcfadden1958axis}, whereas the second one relies on the observation that the velocity field filtered at a scale equal to the integral length scale seems to exhibit uncorrelated zero crossings. Both approaches are able to recover values of $\mathcal{L}$ in several turbulent flows in good agreement with previous established methods \cite{o2004autocorrelation}. Finally, we also analyse the structure of the zero crossings for different turbulent signals by means of Vorono\"{i} tessellations \cite{ferenc2007size,Monchaux2010}.

\section{Methodology}

We analysed measurements taken via hot-wire anemometry (HWA). These measurements, except for those using a passive grid (see table \ref{tab:turbParams}), have been previously published in the literature \cite{dairay2015non,MoraPRF2019}, and span a variety of turbulent flows generated by different mechanisms (see table \ref{tab:turbParams}): downstream of active grids or passive grids, and downstream of the wake of an irregular bluff plate (figure \ref{fig:spk}). 

All grid experiments were conducted in the \textit{Lespinard} wind tunnel in LEGI, a low-turbulence wind tunnel facility with a measurement cross section of 75$\times$ 75 cm$^{2}$, which has been extensively used to conduct experiments under homogeneous isotropic turbulence conditions \cite{MoraPRF2019}.
Measurements were taken 3 m downstream of the grid. The measuring instrument used to record the velocity fluctuations,
was a single Dantec Dynamics 55P01
hot-wire probe, driven by a Dantec StreamLine constant temperature anemometer (CTA) system. The Pt-W wires were 5 $\mu$m in diameter, 3 mm long, with a sensing length of 1.25 mm. Acquisitions were made for 300s at 25kHz and 50 kHz. For all measurements reported here, the Kolmogorov frequency was always smaller than half our sampling frequency.

The wake experiments were conducted in the $3\times3$ wind tunnel at Imperial College London, using the same HWA system as in the grid experiments. Measurements were taken at the centreline at the streamwise distances $D=15$ and $D=50$ from a plate with a characteristic length $D=\sqrt{\cal{A}}=64$mm (with $\cal{A}$ being the frontal area of the plate).

\subsection{Zero crossings computation}
\label{sc:zrcom}

\begin{figure}
	
	\begin{center}
		\begin{subfigure}[t]{0.48\textwidth}
			\includegraphics[scale=0.3]{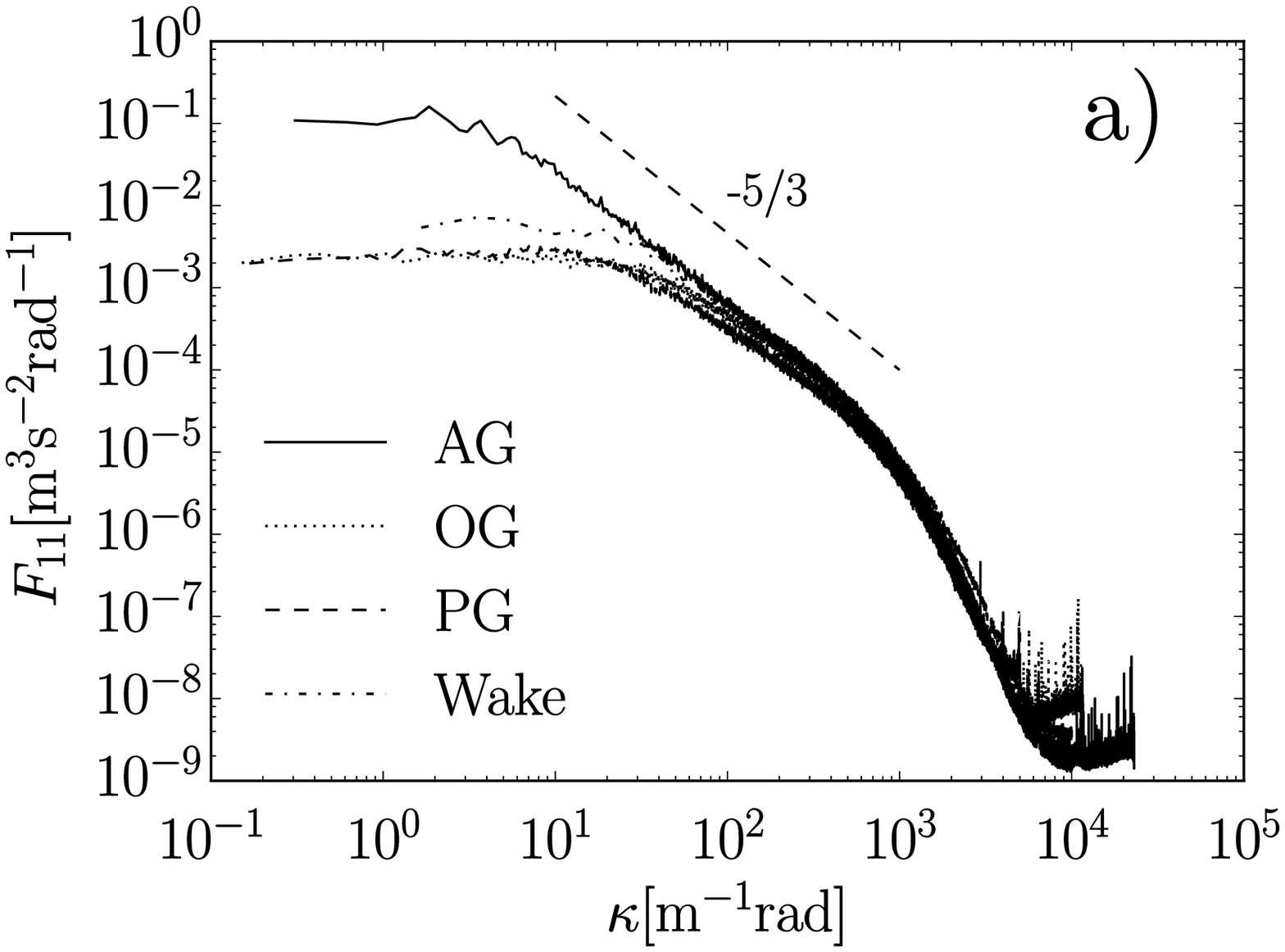}
			\caption{\label{fig:spk}}
		\end{subfigure}
		~
		\begin{subfigure}[t]{0.48\textwidth}
			\includegraphics[scale=0.3]{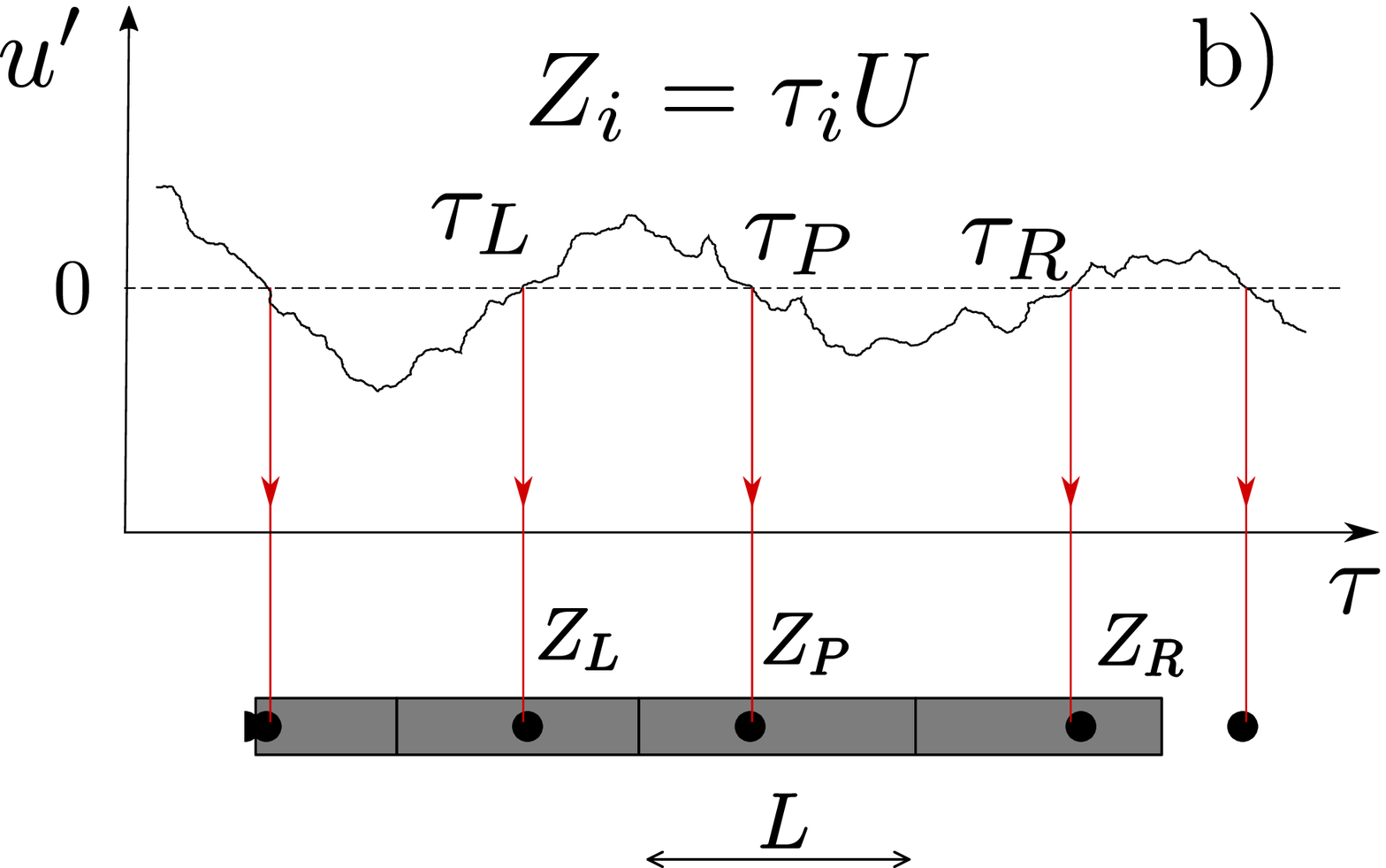}
			\caption{\label{fig:vsk}}
		\end{subfigure}
	\end{center}
	
	\vspace{-0.5cm}
	
	\caption{a) Longitudinal energy density spectra ($F_{11}$) for the data found in table \ref{tab:turbParams}. b) Zero crossings and 1D Vorono\"{i} tessellation illustration. For a given zero crossing position $Z_P$ with left, and right neighbour crossings $Z_L$, and $Z_R$
		respectively, the length of the Vorono\"{i} cell (centered on $Z_P$) is given by $L=\vert Z_R-Z_L \vert /2$.} 
\end{figure}

For this study, we employed a Reynolds decomposition for the streamwise velocity ($u= U+ u^\prime(\tau)$ ), to extract the eulerian fluctuating velocity $u^\prime$. We then computed the fluctuating velocity $u^\prime(\tau)$ zero crossings, i.e., the set of times $\tau^c_i$ for which $u^\prime(\tau^c_i)=0$ (see top of figure \ref{fig:vsk}), and translated this list of zero crossings from time into space by assuming the Taylor hypothesis $Z_i=\tau^c_iU$ (see figure \ref{fig:vsk}). It was verified that all measurements had enough temporal resolution to estimate $\lambda$ \cite{MoraPRF2019} via the zero crossings. A common procedure to verify the latter \cite{sreenivasan1983zero,mazellier2008turbulence} is as follows:

\begin{enumerate}
\item Take the acquired fluctuating velocity signal, and low pass filter it (with a high order filter, e.g. a fifth order butterworth filter) at different sizes $\eta_C=2\pi/\kappa$ (where $\kappa$ is the wave number), given the use of the Taylor hypothesis this is equivalent to  filter at different \textit{frequencies}.
\item Compute the signal zero crossings, and their number density ($n_s=$number of zeros/duration of the signal) at  at each filter size (\textit{frequency}). 
\item If a plateau of $n_s$ is present for filter scales smaller (\textit{larger}) than a certain scale (\textit{frequency}) $\eta_C^\star$ (not to be confused with the Kolmogorov length scale $\eta$), the value of $n_s$ is properly resolved. One could then estimate $\lambda$ via $ n_s^{-1}\vert_\star=\pi C \lambda$, with $C$ being a constant in the order of unity which accounts for the non-gaussianity of the signal \cite{mazellier2008turbulence}.
\end{enumerate}



\begin{table}
	\begin{center}
		
		{\begin{tabular}{ccccc}
				Parameter & OG & AG& PG & Wake \\
				$U_\infty~(m/s)$ & 4.4--17.0 & 1.8--6.8& 1.5--18.0 & 10.0\\
				$\sigma_u/ U ~(\%)$ & 2.0--10.0 & 12.5--15.0& 3.0-2.50& 3.0-8.0\\
				$\lambda~(mm)$ & 8.0--3.0& 16.0--9.0& 9.0--4.0& 10--5.4\\
				$Re_\lambda$ & 50--200 & 200--731& 30--130& 200--300\\
				$\eta~(\mu m)$ & 400--100& 500--180& 918-191& 340--163 \\
				$\mathcal{L}~(cm)$ & 1.0--3.0 & 5.0--11.0&1.7--2.8& 4--6.2 \\
			\end{tabular}}
		\end{center}
		
		\caption{Typical turbulence parameters range for the open (OG), active (AG) \cite{MoraPRF2019}, and  passive (PG) grids. AG refers to the active grid being operated in a random mode while $OG$ to the same grid, completely open and static. We also employed records from the wake of an object \cite{dairay2015non}.  In the table: inlet velocity $U_\infty$, turbulence intensity $\sigma_u/U$ (with $U$ the mean local velocity), Taylor micro-scale $\lambda$, Reynolds number based on the Taylor micro-scale $Re_\lambda=\lambda\sigma_u/\nu$ ($\nu$ being the kinematic viscosity of the flow), Kolmogorov length scale $\eta=(\nu^3/\varepsilon)^{1/4}$ and the streamwise integral length scale $\mathcal{L}$ (obtained via $\rho(r)$). \label{tab:turbParams}}	
	\end{table}
	
	\subsection{Variance of zero crossing successive intervals and $\mathcal{L}$} \label{sc:intL}
	The seminal work of McFadden \cite{mcfadden1958axis} was the first to derive for gaussian processes closed expressions to compute the variance of the interval distance between two successive zero crossings ($\Delta Z=Z_{i+1}-Z_{i}$) under two analytically tractable conditions: intervals between zeros are statistically independent, or intervals between zeros make a Markov chain. For the former, statistically independent case, the expression for the variance goes as,
	
	\begin{equation}
	\mathrm{Var}(\Delta Z)=2\langle \Delta Z\rangle\int_0^\infty \frac{2}{\pi}\mathrm{arcsin}(\rho(\tau))d\tau.
	\label{eq:varo}
	\end{equation}
	
	The latter expression, and assuming Taylor hypothesis, provides an estimate for $\mathcal{L}$ if the assumption of independent zero crossing intervals approximately holds for turbulent signals;
	
	\begin{equation}
	\frac{\mathrm{Var}(\Delta Z)}{2\langle \Delta Z\rangle}=\int_0^\infty \frac{2}{\pi}\mathrm{arcsin}(\rho(\tau))d\tau=\int_0^\infty \frac{2}{\pi}\Big(\rho(\tau)+\frac{\rho(\tau)^3}{6}+\frac{\rho(\tau)^5}{40}+\cdots \Big) \leq \int_0^\infty \rho(\tau)d\tau.
	\label{eq:ff}
	\end{equation}
	
	By truncating the integral up to the first term, one obtains a relation between the Fano factor \cite{smith2008fluctuations}, and the integral time ($\mathcal{T}$) and length scales:
	
	\begin{equation}
	\frac{\pi}{4}\frac{\mathrm{Var}(\Delta Z)}{\langle \Delta Z\rangle}\approx \int_0^\infty \rho(\tau)d\tau=\mathcal{T}=\mathcal{L}/U.
	\label{eq:vartr}
	\end{equation}
	
	\subsection{Successive zero crossings and 1D Vorono\"{i} tessellation}
	\label{sec:vorozc}
	Interestingly, if McFadden's assumption of statistically independent intervals holds, the variance of the length between two successive intervals \cite{bendat2011random} could be written as,
	\begin{equation}
	\mathrm{Var}(\Delta Z_1+\Delta Z_2)=\mathrm{Var}(\Delta Z_1)+\mathrm{Var}(\Delta Z_2)+\cancelto{0}{\mathrm{Cov}(\Delta Z_1,\Delta Z_2)}.
	\label{eq:idpv}
	\end{equation}
	
	Thus, an easy way to analyse the implications of this assumption (and its deviations) could be using 1D Vorono\"{i} tessellations \cite{ferenc2007size}. These two frameworks (Vorono\"{i} tessellations and `raw' inter-crossing distances) are related by their respective definitions. For instance, if a random variable $\Delta Z$ represents the distance between successive crossings, the respective Vorono\"{i} cell length  $L$ (also a random variable) is given by $L= 1/2(\Delta Z_L+\Delta Z_R)$, where $\Delta Z_L=\vert Z_P-Z_L \vert$ and $\Delta Z_R=\vert Z_R-Z_P \vert$ are the crossing lengths (random variables) at the left, and at the right of the crossing $Z_P$ (see figure \ref{fig:vsk}). From these definitions; $\langle L\rangle =\langle \Delta Z\rangle=n^{-1}_s$.
	
	Then, if the covariance between $\Delta Z_L$ and $\Delta Z_R$  is very weak, the variance of the ensemble of normalized Vorono\"{i} cells ($\mathcal{V}=L/\langle L \rangle $) is half the variance of $\Delta Z/\langle \Delta Z\rangle^2$:
	
	\begin{equation}
	\frac{1}{2}\frac{\mathrm{Var}(\Delta Z)}{\langle \Delta Z\rangle^2}= \frac{\mathrm{Var}(L)}{\langle L\rangle^2}=	\sigma_\mathcal{V}^2.
	\label{eq:varrpp}
	\end{equation}
	
	The latter expression is easily verified for a random Poisson process (RPP), for which $\langle \Delta Z\rangle=1$, $\mathrm{Var}(\Delta Z)=1$, and its respective Vorono\"{i} normalised cell variance is $\mathrm{Var}(L)/\langle L\rangle^2=1/2$ \cite{ferenc2007size}. We will refer to the standard deviation of this RPP process as $\sigma_{RPP}=\sqrt{1/2}$, and to the respective standard deviation coming from our Vorono\"i analysis of turbulent signals as $\sigma_\mathcal{V}$. 
	
	\section{Results}
	
	\subsection{Estimation of $\mathcal{L}$ via the McFadden equation}
	
	We compute the values of $\mathrm{Var}(\Delta Z)$ and $\langle \Delta Z\rangle$ for all datasets and examine the accuracy of  McFadden's equation (equation \ref{eq:vartr}). To estimate $\mathrm{Var}(\Delta Z)$ and $\langle \Delta Z\rangle$, we follow a standard procedure found in the literature \cite{sreenivasan1983zero,mazellier2008turbulence}, and described in section \ref{sc:zrcom}: we low-pass filter $u^\prime$ with a range of filter sizes $\eta_C$, and compute the signal zero crossings and zero crossings density at each filter size. The presence of a plateau (not reported here, but present for all our datatets) for small values of $\eta_C$ show that $n_s$ is well resolved, and therefore $\lambda$ can be computed as $ n^{-1}_s= \langle \Delta Z\rangle \sim \pi \lambda $. Resolving $\lambda$ gives credence to the use of McFadden's equation with our datasets values. 
	
\begin{figure}
		\centering
		\begin{center}
			\begin{subfigure}[t]{0.48\textwidth}
				\includegraphics[scale=0.3]{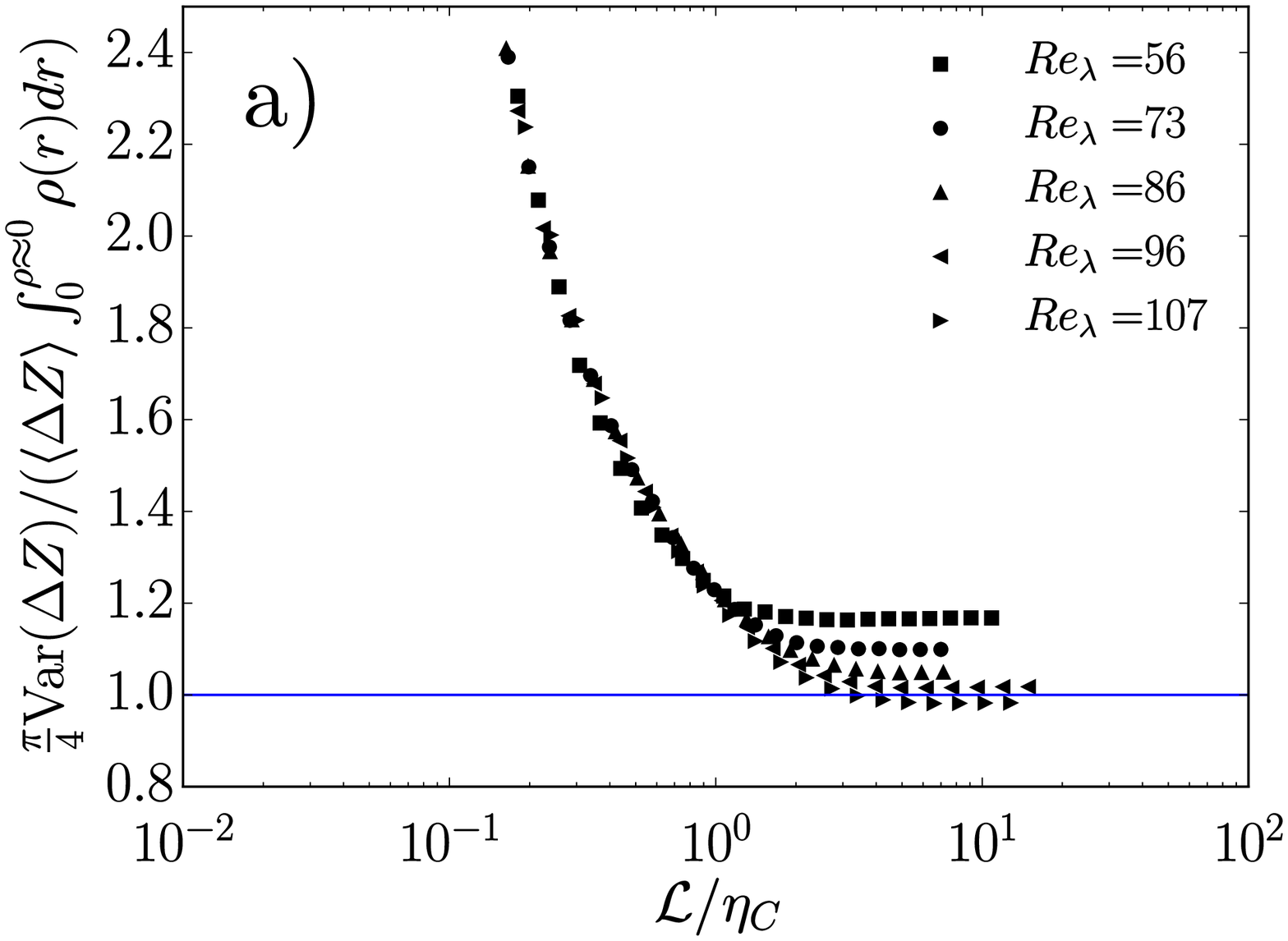}
				\caption{\label{fig:LINXOG}}
			\end{subfigure}
			~
			\begin{subfigure}[t]{0.48\textwidth}
				\includegraphics[scale=0.3]{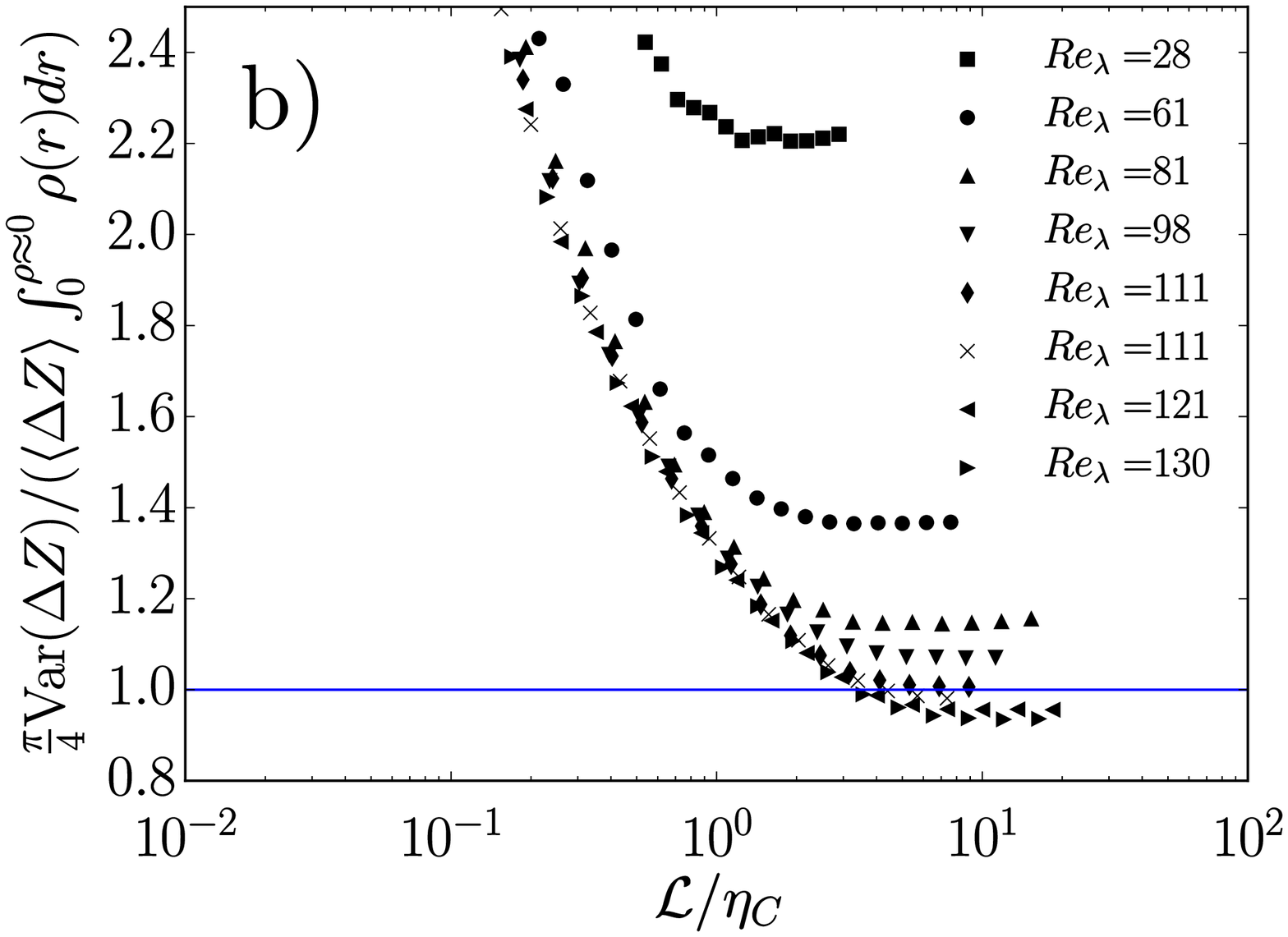}
				\caption{\label{fig:LINXPG}}
			\end{subfigure}
			
		\end{center}
				\vspace{-1cm}
		\begin{center}		
			
			\begin{subfigure}[t]{0.48\textwidth}
				\includegraphics[scale=0.3]{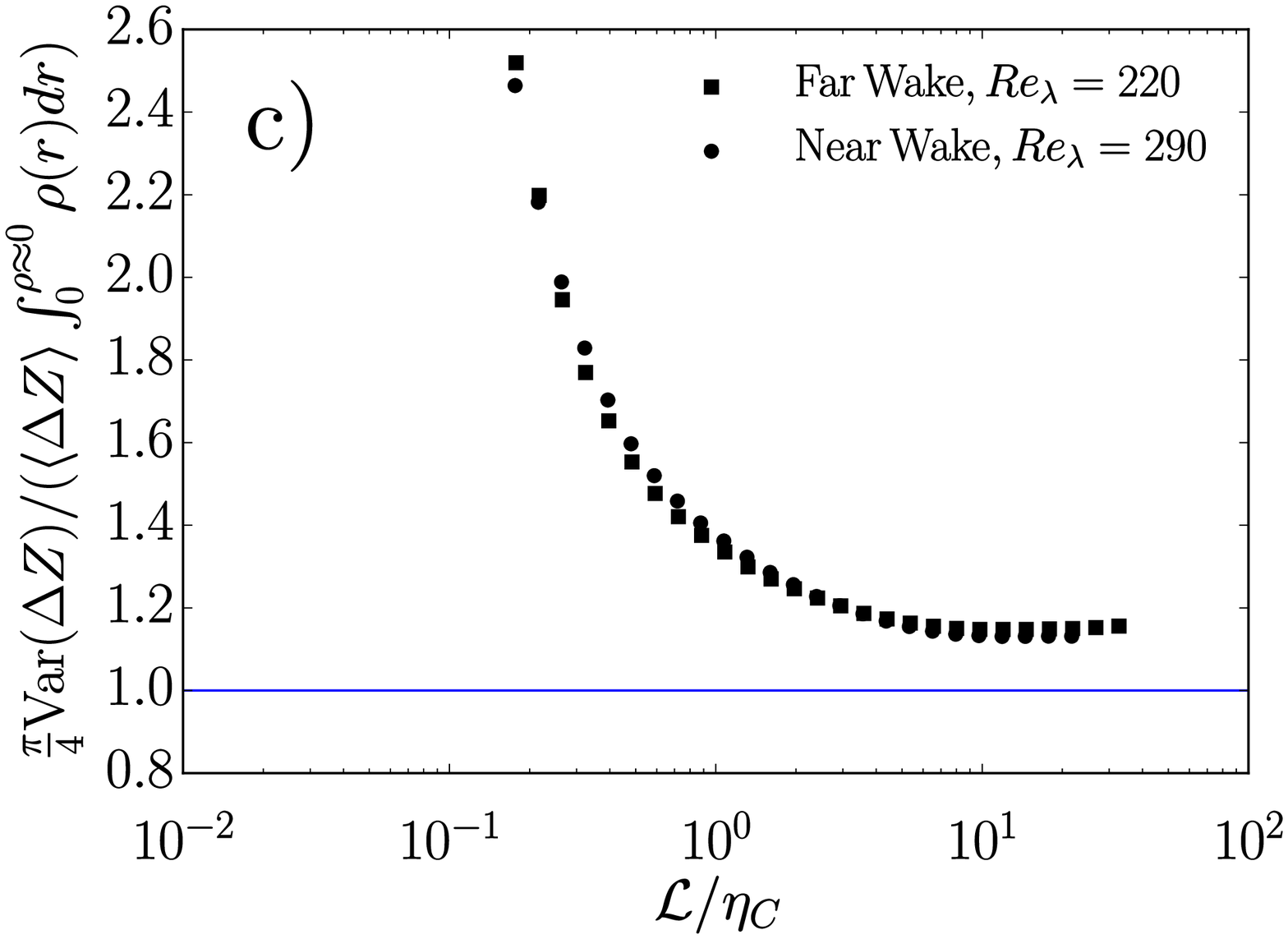}
				\caption{\label{fig:LINXWK}}
			\end{subfigure}
			~
			\begin{subfigure}[t]{0.48\textwidth}
				\includegraphics[scale=0.3]{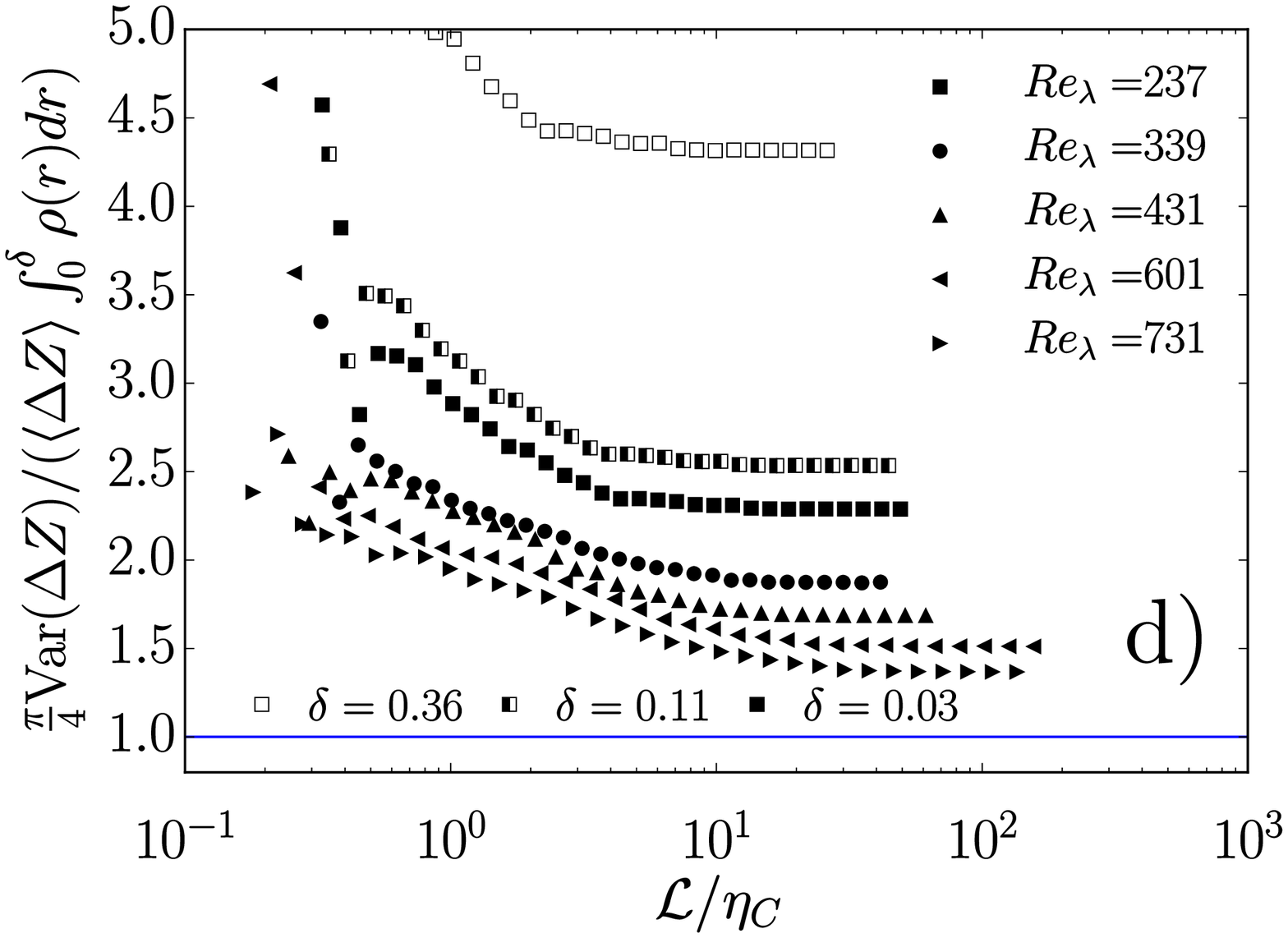}
				\caption{\label{fig:LINXAG}}
			\end{subfigure}

		\end{center}
		\vspace{-0.5cm}
		\caption{Estimation of $\mathcal{L}$ via equation \ref{eq:vartr}. a) OG. b) PG. c) Wakes. d) AG. For the AG datasets \cite{MoraPRF2019}, the autocorrelation integrals were estimated by the method of Puga and La Rue \cite{puga2017normalized}. We also show the sensitivity of this parameter to different values of $\delta$ for the lowest value of $Re_\lambda$ ($\square$).
			\label{fig:inauto}}
	\end{figure}

	To check the validity of McFadden's equation, we apply it to all velocity signals and for all filter scales (see figure \ref{fig:inauto}). The equation provides indeed an acceptable estimation of $\mathcal{L}$ for large values of $Re_\lambda$ when the value at the plateau is compared to the traditional method of integrating up to the first zero of the autocorrelation. 
	
	Our method, however, has a residual dependency on $Re_\lambda$ coming from variance value at the plateau, and as expected from previous analyses \cite{mazellier2008turbulence}.  On the other hand, for the AG data (figure \ref{fig:LINXAG}), our results suggest that there was indeed a underestimation of the original values of $\mathcal{L}=\int_0^{r_0} \rho(r)dr$ (already discussed in \cite{MoraPRF2019}), and such underestimation strongly depends on the value of $\delta$ selected; as $\delta$ increases $\mathcal{L}$ decreases.

\begin{figure}

			\begin{center}

				\begin{subfigure}[t]{0.48\textwidth}
					\includegraphics[scale=0.3]{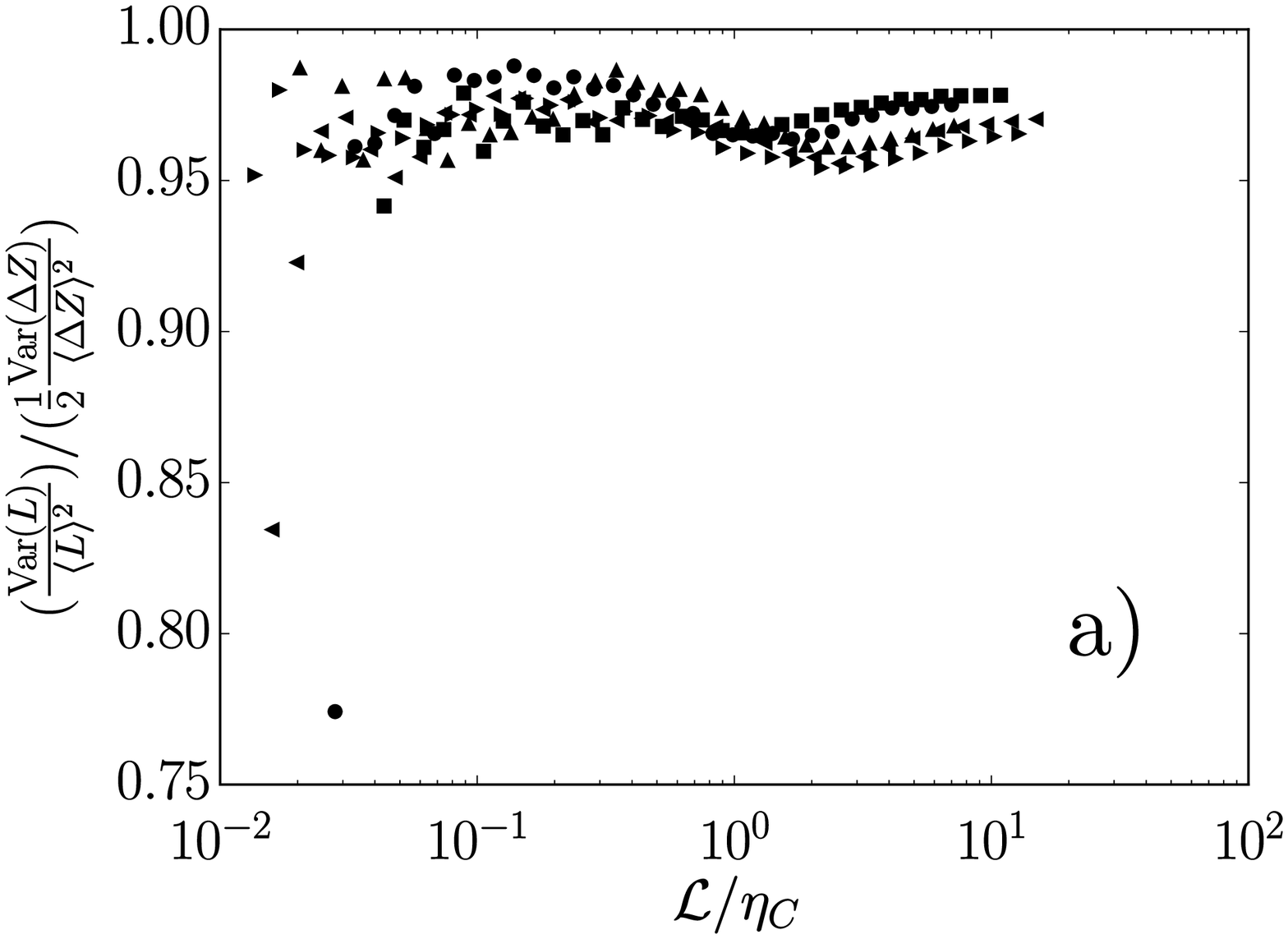}
					\caption{ \label{fig:STDROG}}
				\end{subfigure}
				~
				\begin{subfigure}[t]{0.48\textwidth}
					\includegraphics[scale=0.3]{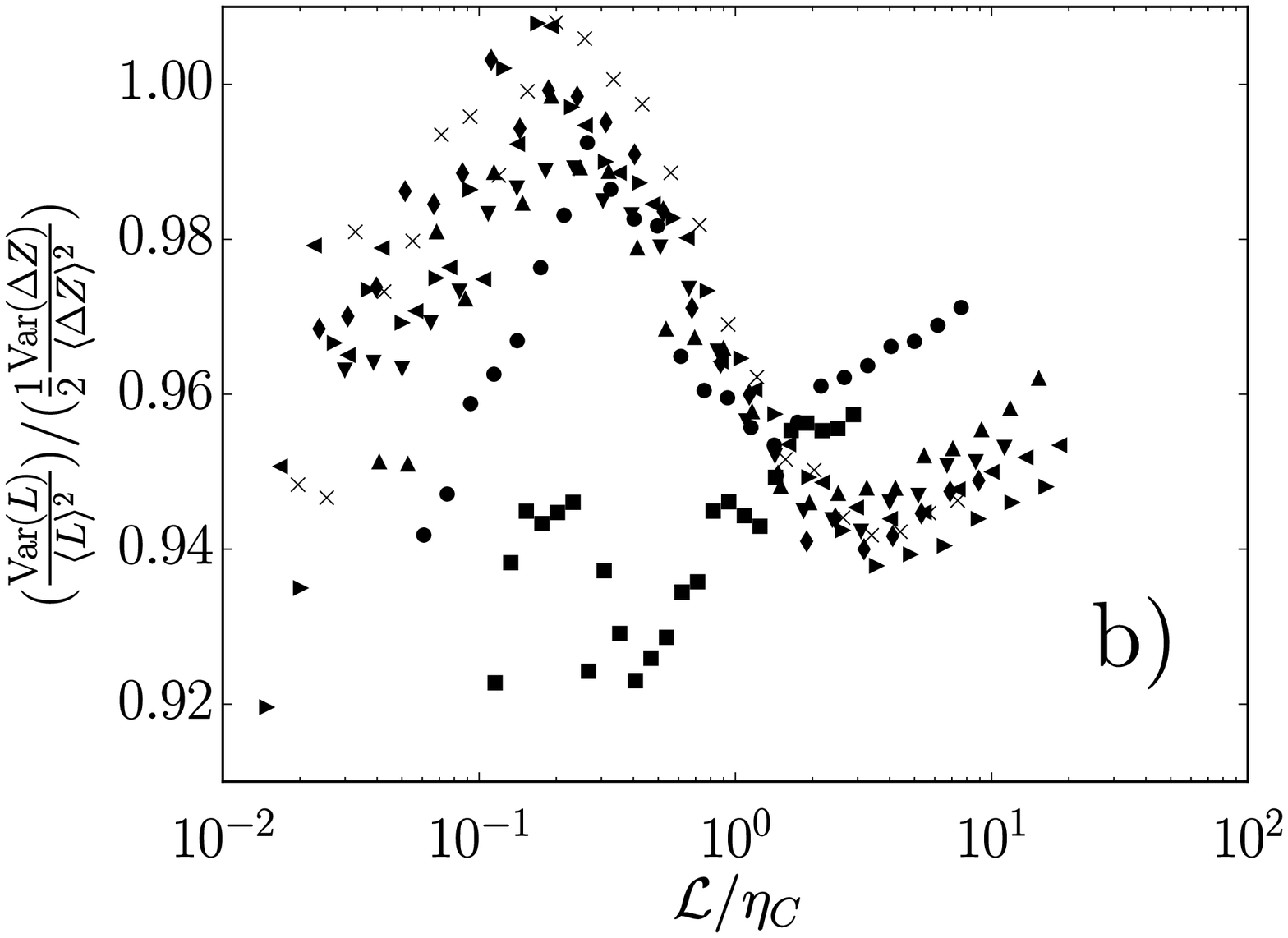}
					\caption{\label{fig:STDpG}}
				\end{subfigure}
								
			\end{center}
					\vspace{-1cm}
			\begin{center}		
				\begin{subfigure}[t]{0.48\textwidth}
					\includegraphics[scale=0.3]{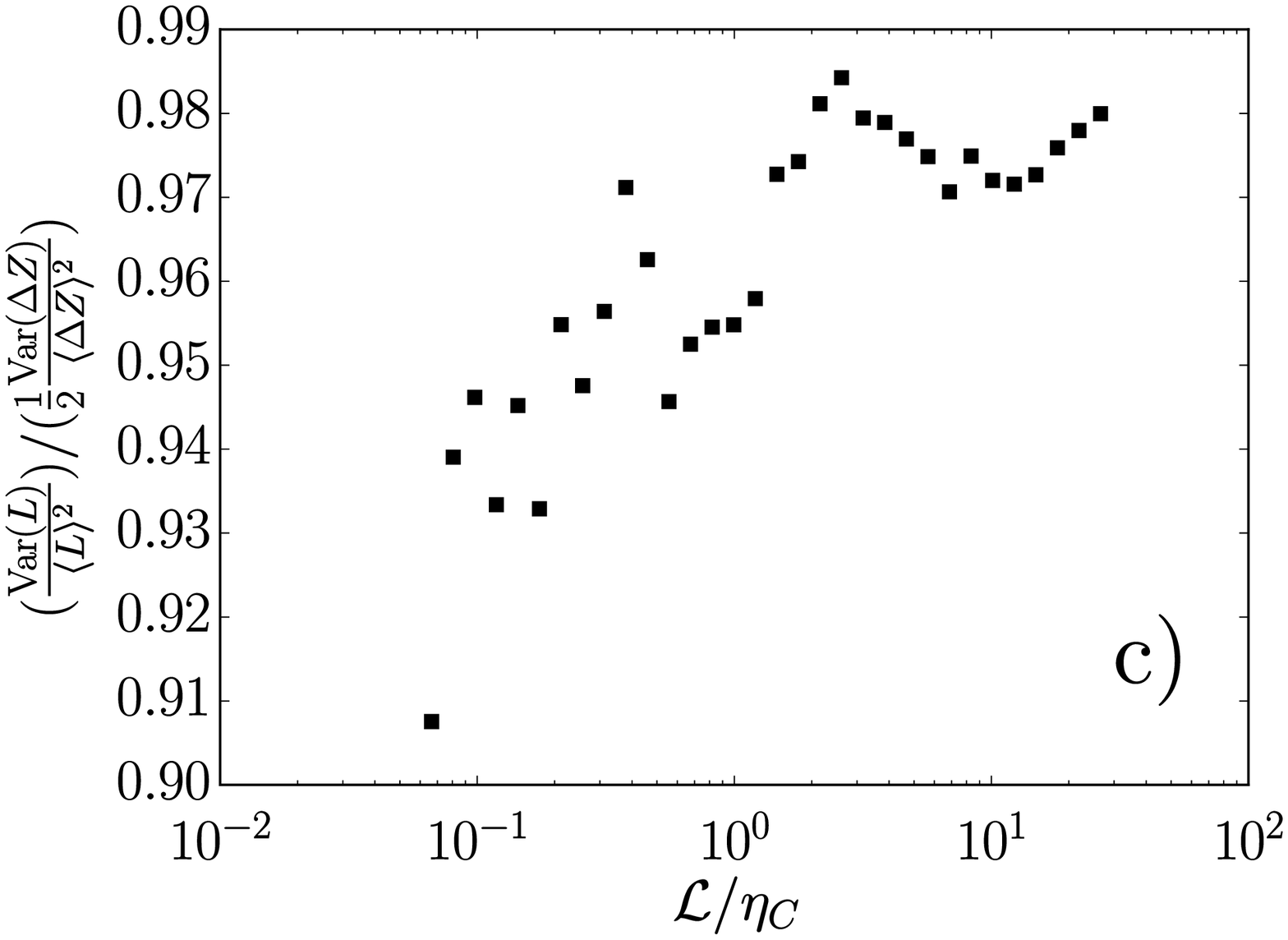}
					\caption{\label{fig:STRDWK}}
				\end{subfigure}
				~
				\begin{subfigure}[t]{0.48\textwidth}
					\includegraphics[scale=0.3]{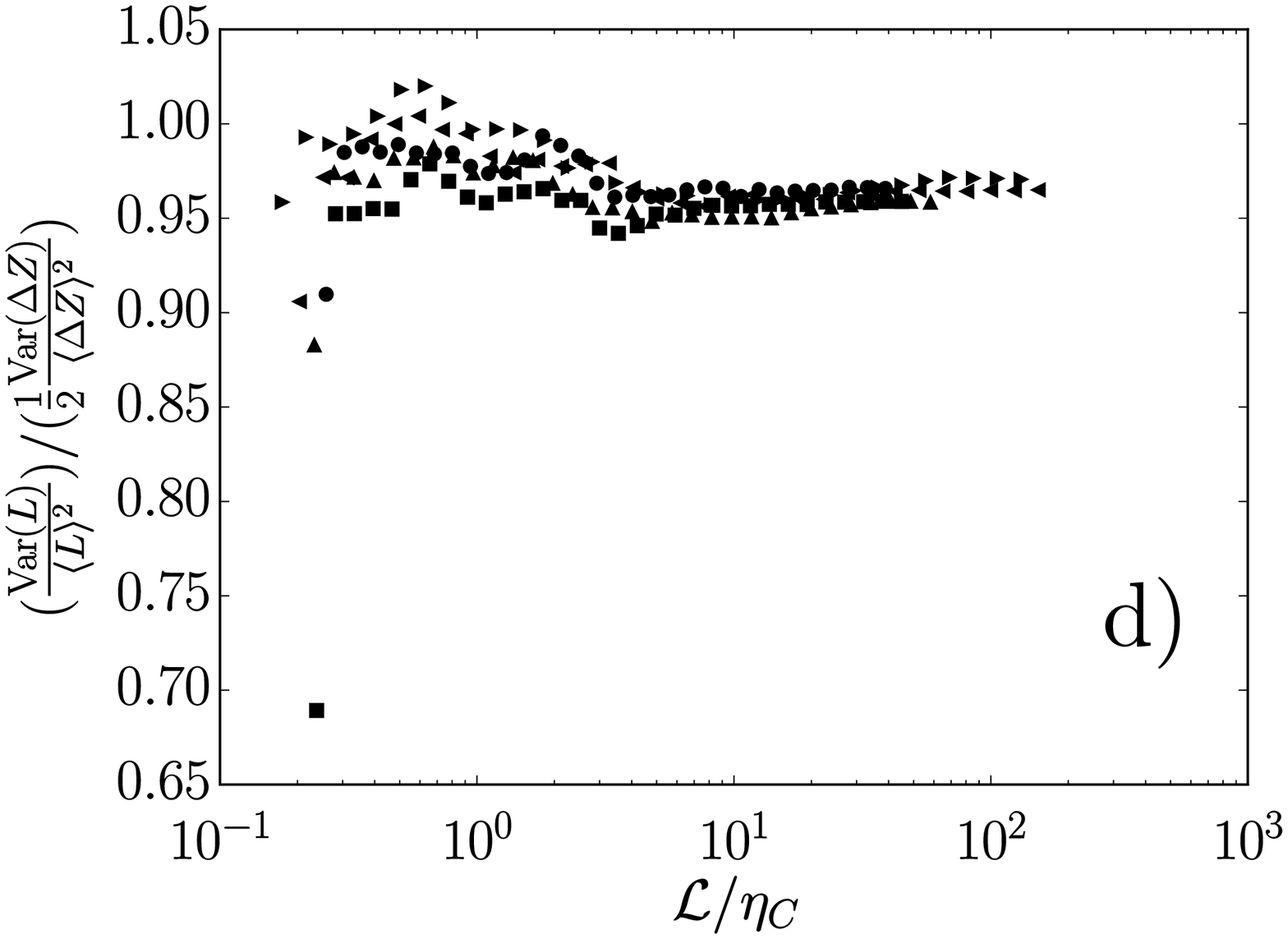}
					\caption{ \label{fig:STDRAG}}
				\end{subfigure}
			\end{center}
						\vspace{-0.5cm}
		\caption{Ratio between the variance of Vorono\"{i} cells $\mathrm{Var}(L)=\langle (L-\langle L\rangle)^2\rangle$, and the respective one of  the interval distance between two successive zero crossings; $\mathrm{Var}(\Delta Z)=\langle \Delta Z^2\rangle - \langle \Delta Z\rangle^2$. Note that $\langle L \rangle=\langle \Delta Z \rangle $. a) OG. b) PG. c) Wakes. d) AG.}
	\end{figure}

	Moreover, these observations advance that indeed the assumption of independent successive zero crossings could hold to some extent in turbulent signals.  In fact, if equation \ref{eq:idpv} is approximately valid at all filters of interest, then the same follows for equation \ref{eq:varrpp} , which relates the variance of the Vorono\"{i} tessellation $\sigma_\mathcal{V}^2$ applied to zero crossings with the variance of the interval distance between two successive zero crossings $\mathrm{Var}(\Delta Z)$. The latter yields $\mathrm{Var}(L)=\frac{1}{2} \mathrm{Var}(\Delta Z)$. Equation \ref{eq:varrpp} then (see figures  \ref{fig:STDROG} to \ref{fig:STDRAG}) gives: 
	\begin{equation}
	0.90<\sigma_\mathcal{V}^2/\Big(\frac{1}{2}\frac{\mathrm{Var}(\Delta Z)}{\langle \Delta Z\rangle^2}\Big)=\Big(\frac{\mathrm{Var}(L)}{\langle L\rangle^2}\Big)/\Big(\frac{1}{2}\frac{\mathrm{Var}(\Delta Z)}{\langle \Delta Z\rangle^2}\Big)<1.02,
	\end{equation}
	at all filter scales of interest for the data found in Table \ref{tab:turbParams}. This observation supports that the independent interval assumption is approximately valid within $10\%$ error for our datasets.
	
	\subsection{Vorono\"{i} analysis} 
	
	Taking into account the previous section observations, we applied 1D Vorono\"{i} tessellation analysis \cite{ferenc2007size} to the signals' zero crossings (see figure \ref{fig:vsk}) at each filter scale.
	
\begin{figure}	
		\begin{center}
			
			\begin{subfigure}[t]{0.48\textwidth}
				\includegraphics[scale=0.3]{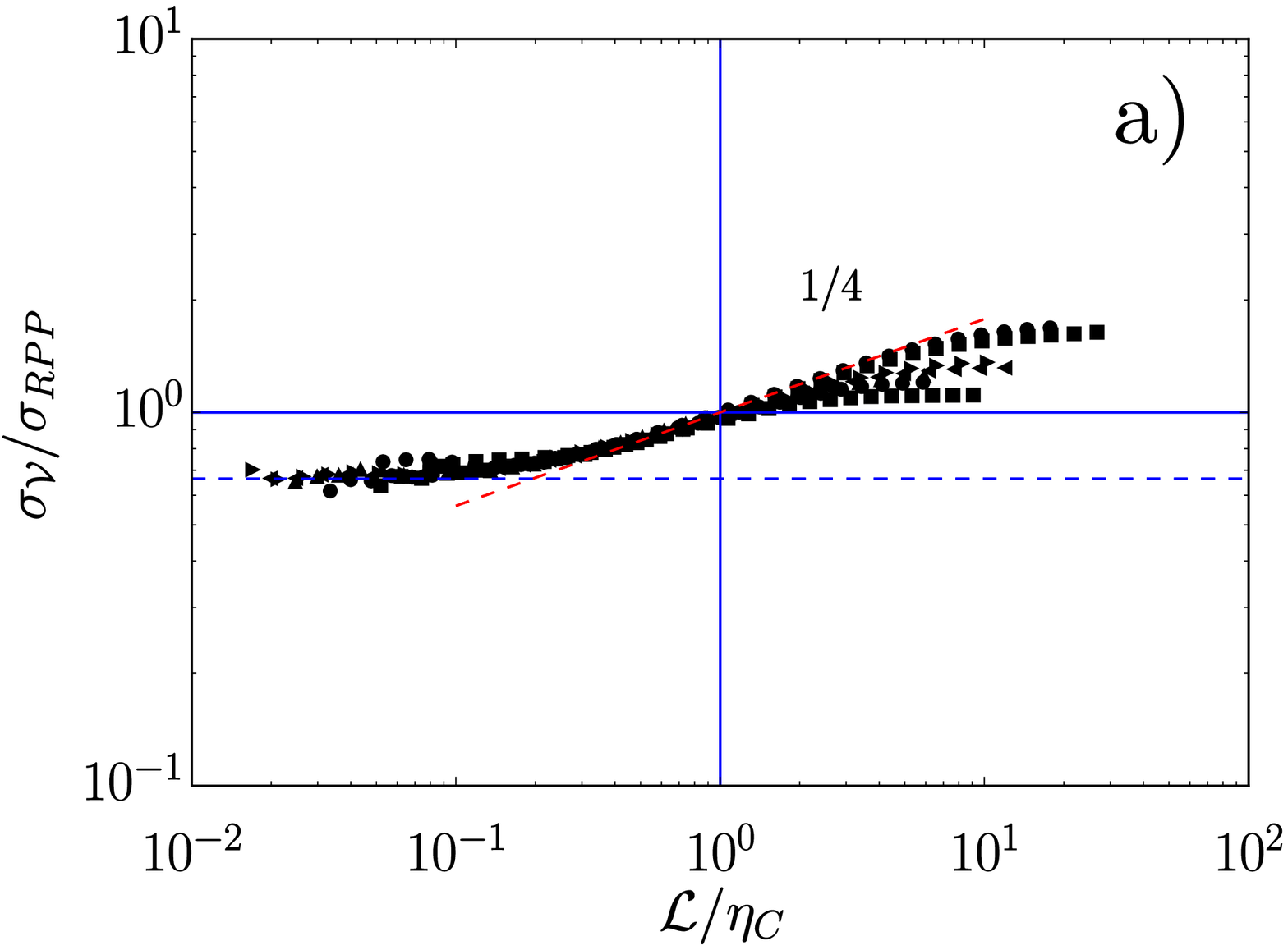}
				\caption{ \label{fig:STDOG}}
			\end{subfigure}
			~
			\begin{subfigure}[t]{0.48\textwidth}
				\includegraphics[scale=0.3]{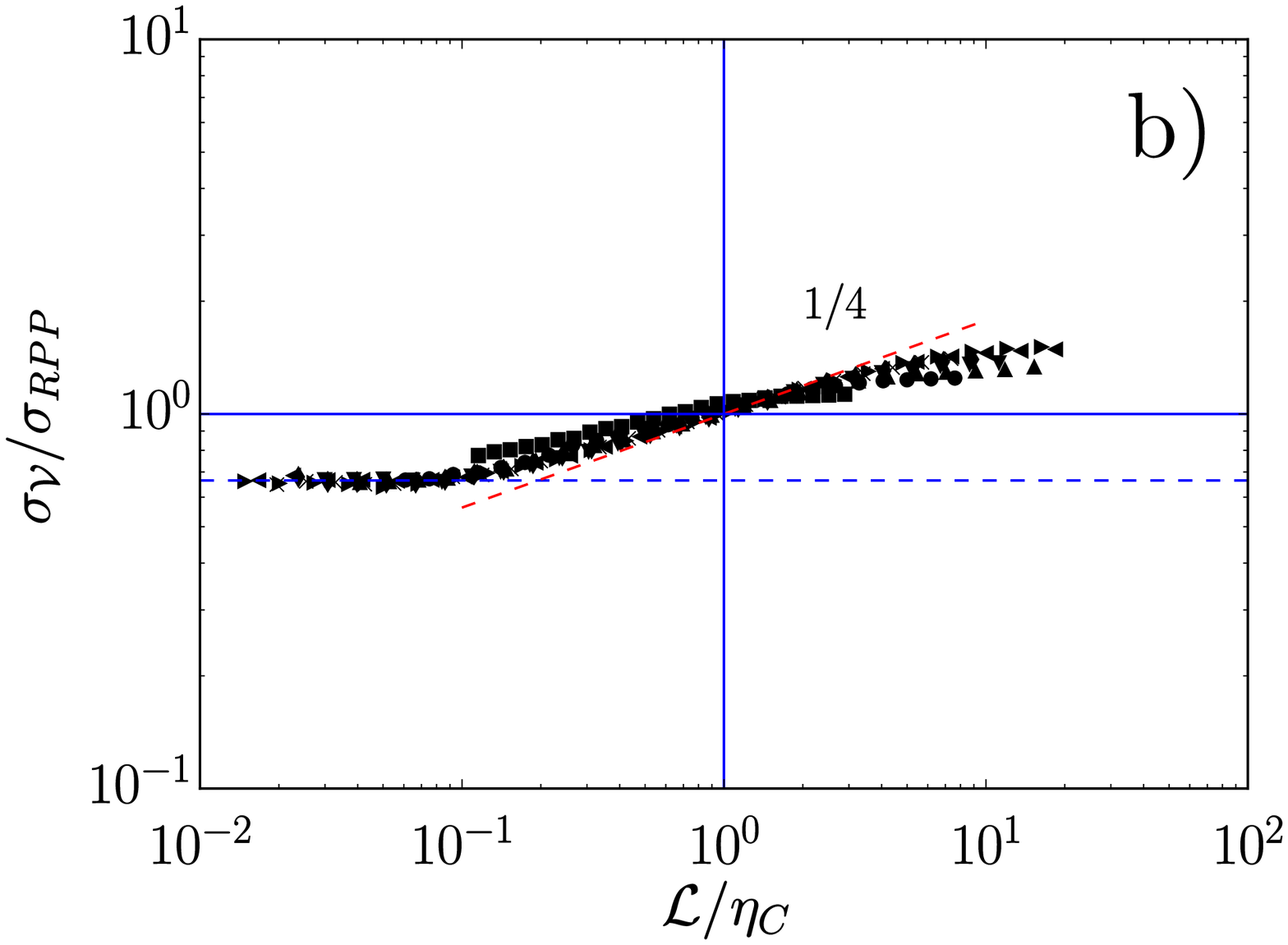}
				\caption{\label{fig:STDPG}}
			\end{subfigure}
			
		\end{center}
		\vspace{-1cm}
		\begin{center}		
			\begin{subfigure}[t]{0.48\textwidth}
				\includegraphics[scale=0.3]{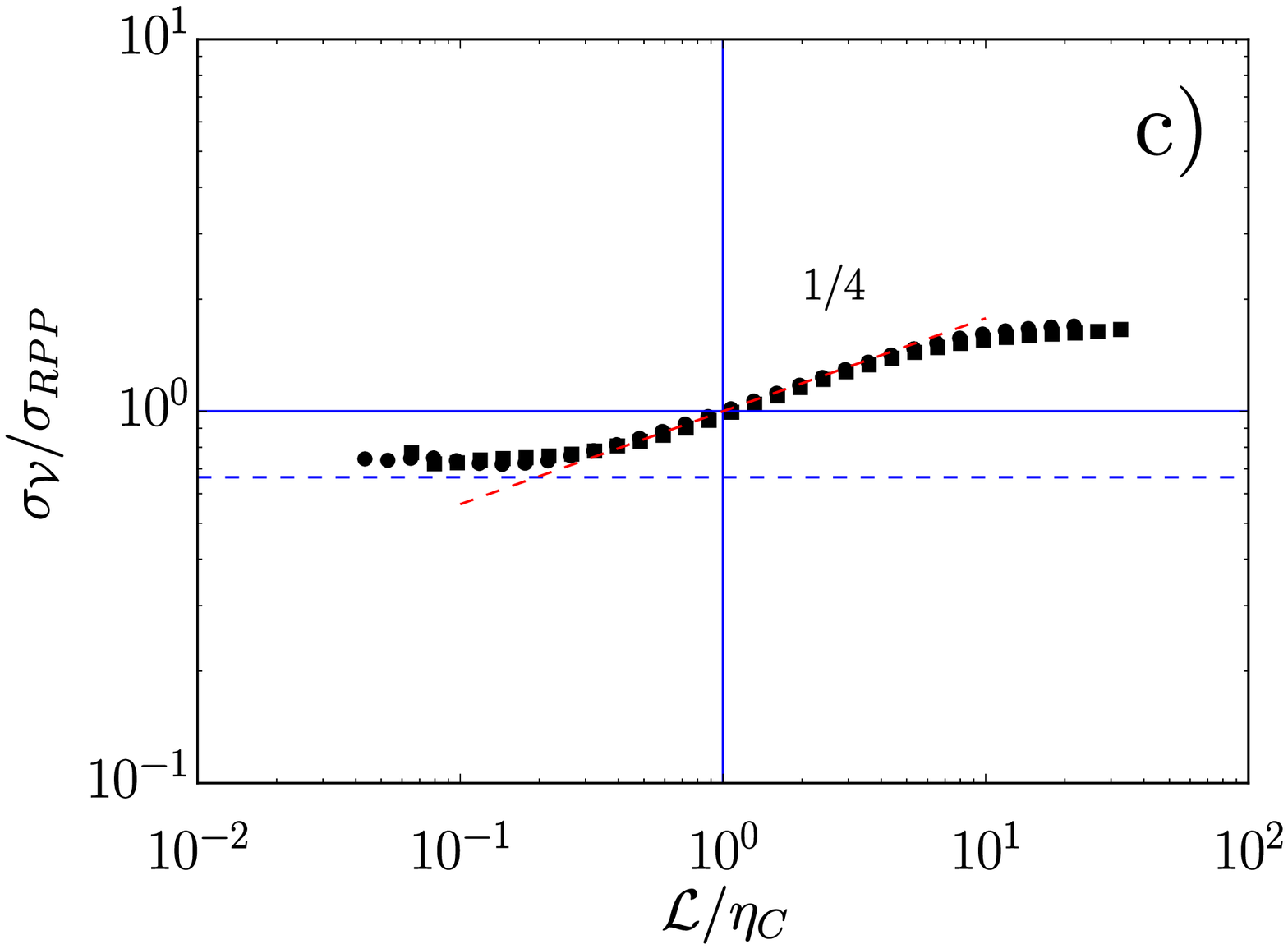}
				\caption{\label{fig:STDWK}}
			\end{subfigure}
			~
			\begin{subfigure}[t]{0.48\textwidth}
				\includegraphics[scale=0.3]{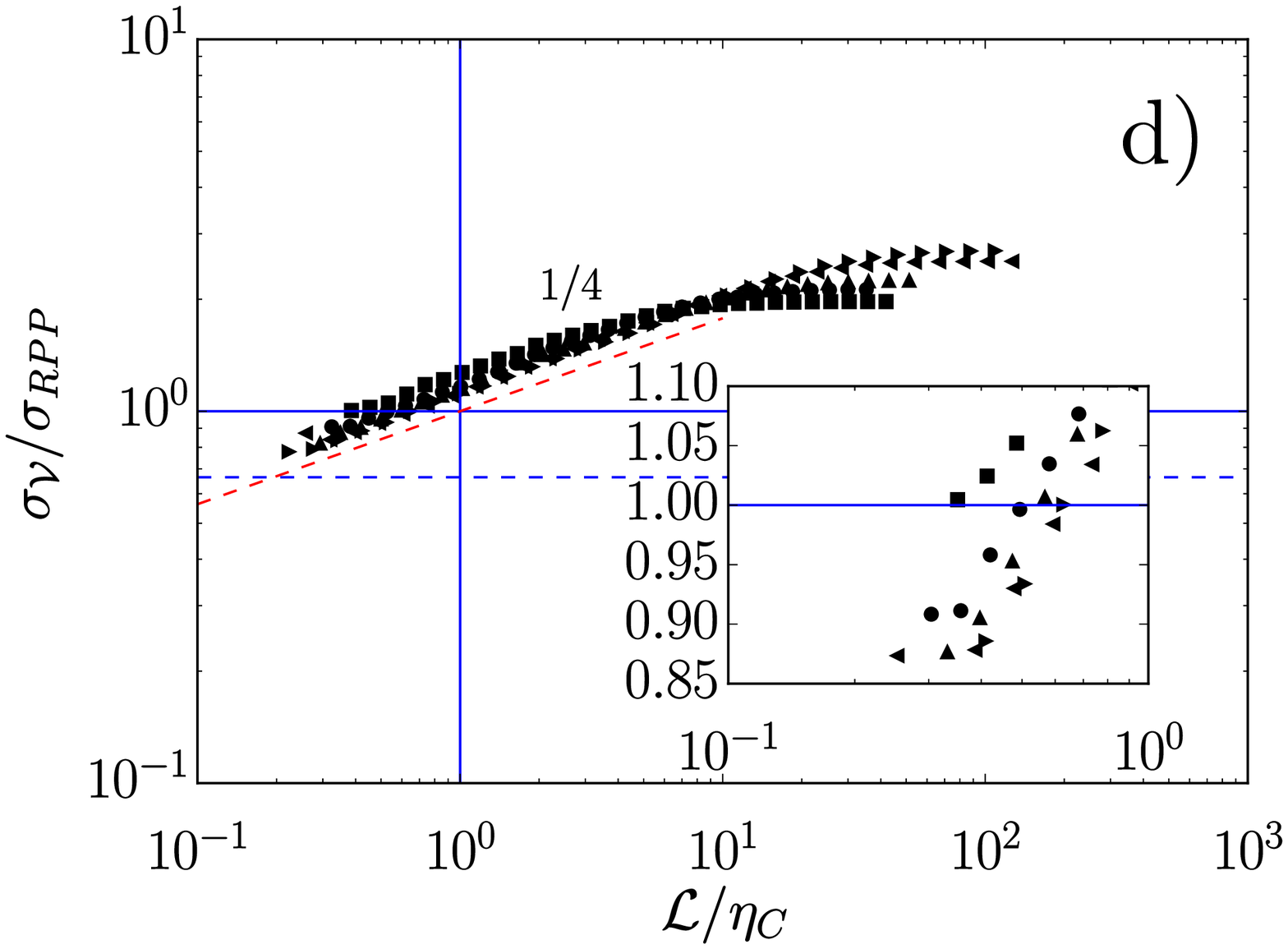}
				\caption{\label{fig:STDAG}}
			\end{subfigure}
			
		\end{center}
		\vspace{-0.5cm}
		\caption{Evolution of $\sigma_\mathcal{V}$ against the filter scale. Once more, in all cases $\mathcal{L}$ has been obtained from the autocorrelation function. a) OG. b) PG. c) Wakes. d) AG. In d) we are only plotting the AG data for $\delta=0.03$ (see figure \ref{fig:LINXAG}). In the figures the horizontal dashed line correspond to the standard deviation corresponding to a gaussian white noise process \cite{mcfadden1958axis}; $\sigma_\mathcal{V}/\sigma_{RPP}\approx0.67$.}
	\end{figure}

	Our results reveal that $\sigma_\mathcal{V}$ (see figures \ref{fig:STDOG} to \ref{fig:STDAG}) has complex behaviour with the filter scale $\eta_C$ which can be divided in 3 parts. First, a plateau regime at low values of $\kappa_C\sim\eta_C^{-1}$ representative of a flat white noise gaussian spectrum. Second, an intermediate regime where $\sigma_\mathcal{V}$ may attain a power law behaviour with an exponent close to $1/4$ for large values of $Re_\lambda$; despite of its persistence among different datasets, the existence of this intermediate regime remains unexplained and is left for future research as the limited extend of our data does not allows us to unambiguously conclude the accuracy of the exponent. Third, a second plateau consistent with the one found for the zero crossing number density $n^{-1}_s=\langle \Delta Z\rangle$ is found. 
	
	On the contrary, and interestingly, there seems to be a strong correlation between the filter scale at which $\sigma_\mathcal{V}/\sigma_{RPP}\approx1$, and $\mathcal{L}$, i.e., apparently $\sigma_\mathcal{V}/\sigma_{RPP}\approx1$ when $\eta_C=\mathcal{L}$. These results hint that an alternative definition of $\mathcal{L}$ could be: the length-scale at which the zero crossing intervals topology is approximately uncorrelated.
	
\begin{figure}
		\begin{center}
			\begin{subfigure}[t]{0.48\textwidth}
				\includegraphics[scale=0.3]{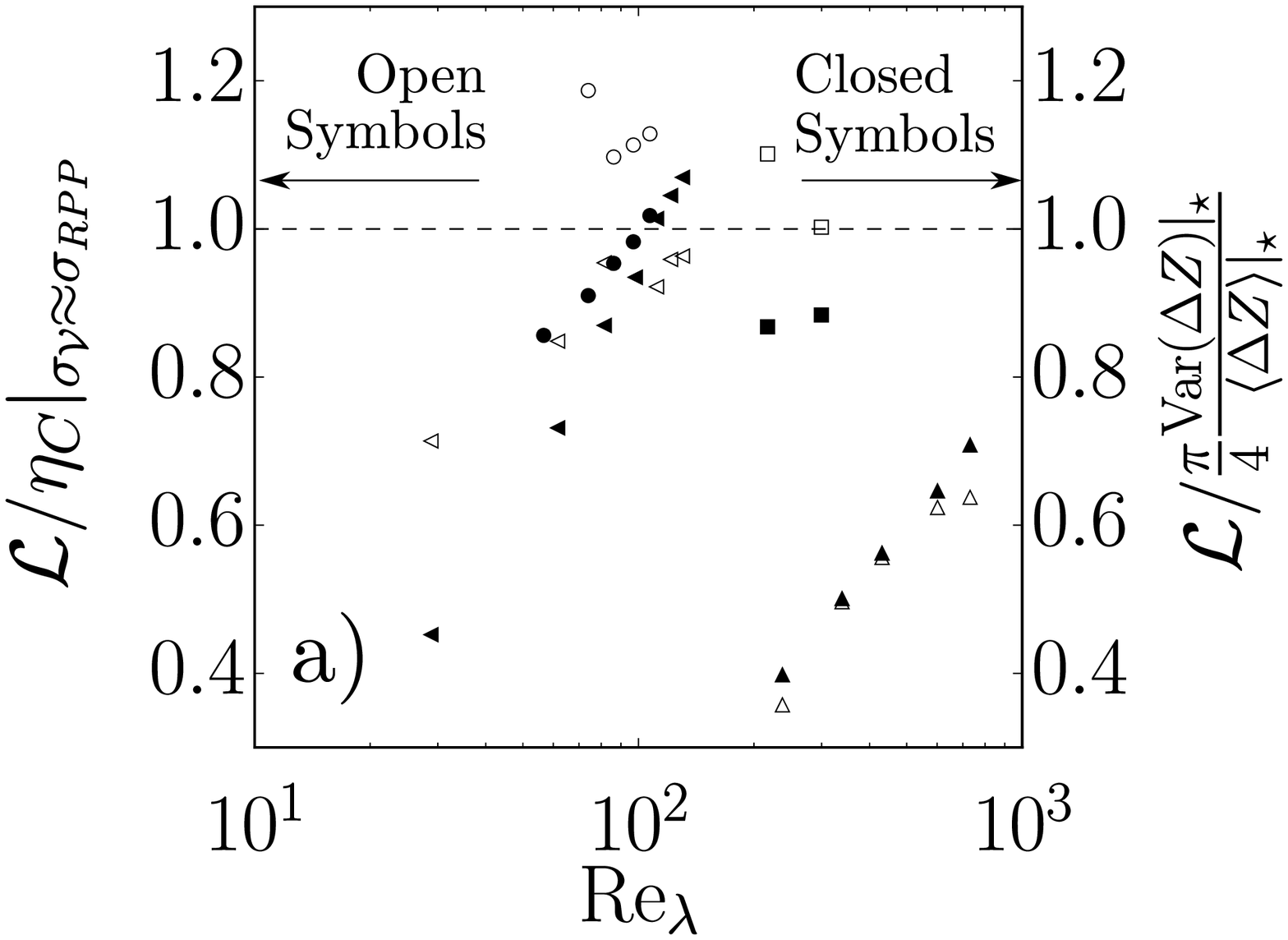}
				\caption{\label{fig:compx}}
			\end{subfigure}
			~
			\begin{subfigure}[t]{0.48\textwidth}
				\includegraphics[scale=0.3]{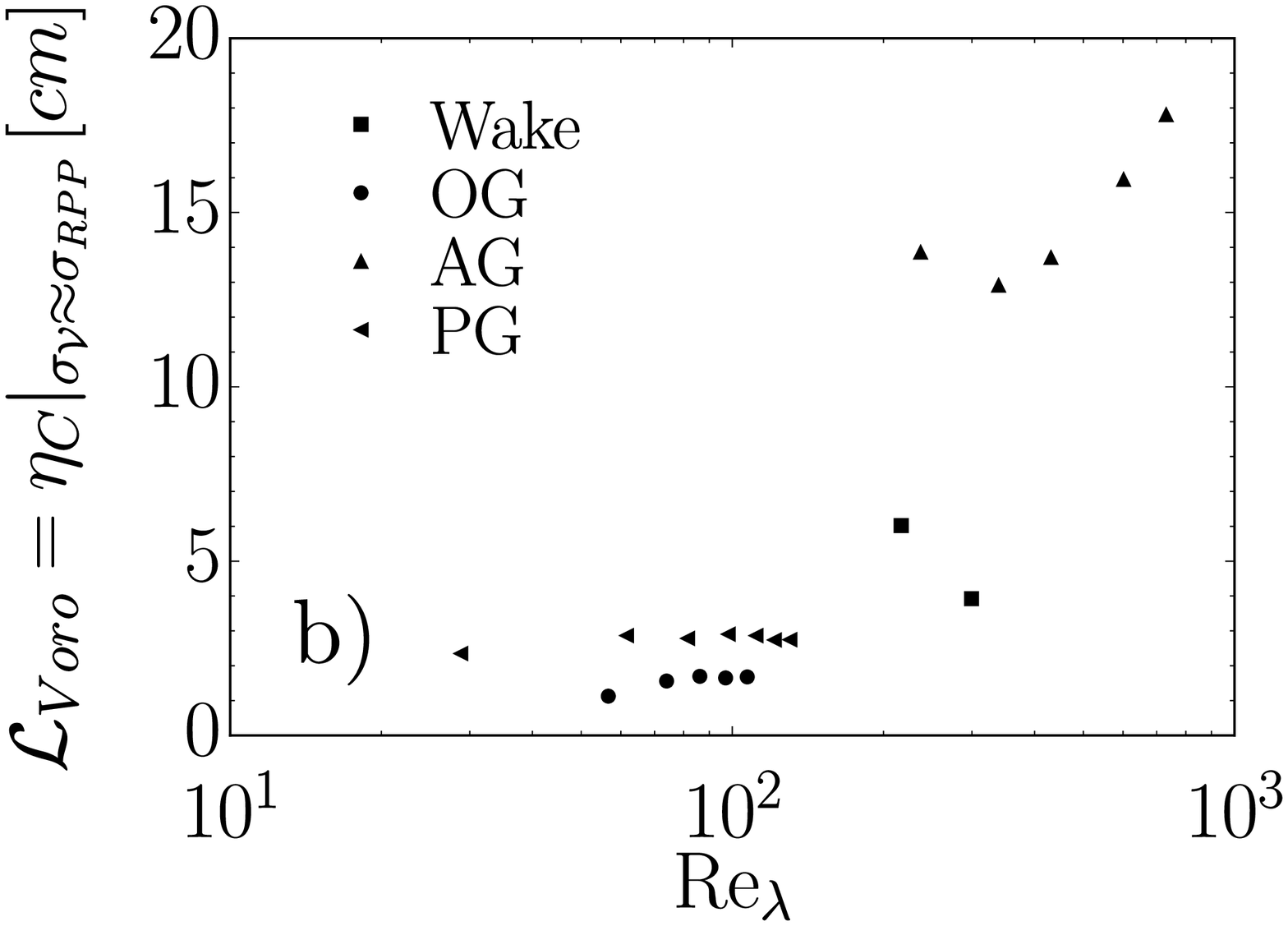}
				\caption{\label{fig:compx2}}
			\end{subfigure}
					\end{center}
								\vspace{-1cm}
	\begin{center}	
			\begin{subfigure}[t]{0.48\textwidth}
				\includegraphics[scale=0.3]{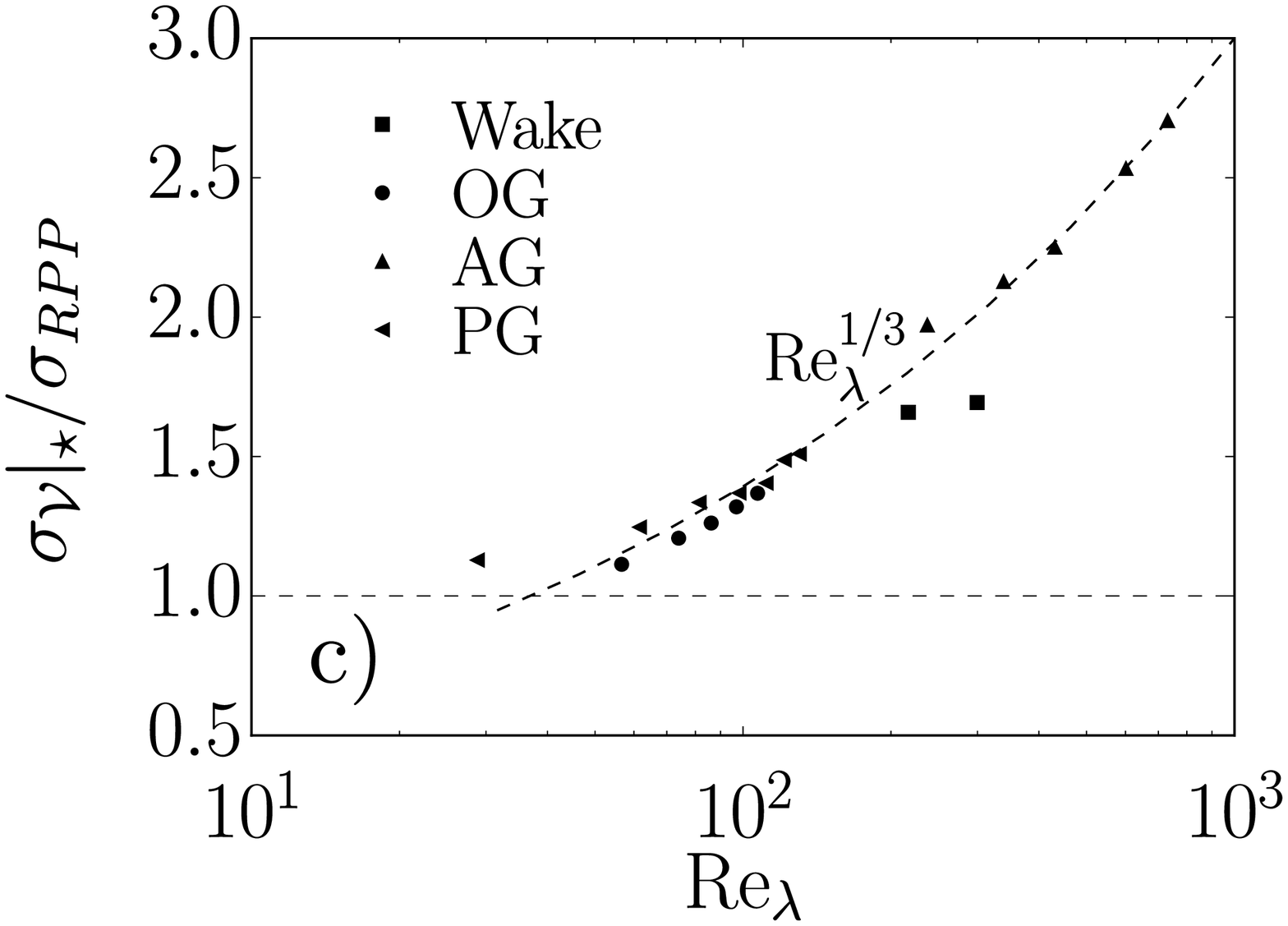}
				\caption{\label{fig:stpx}}
			\end{subfigure}
		\end{center}
		\vspace{-0.5cm}
		\caption{Comparison between the two methods to estimate $\mathcal{L}$.  a) Estimation of $\mathcal{L}$ from $\mathcal{L}_{Voro}=\eta_C$ for which $\sigma_\mathcal{V}\approx \sigma_{RPP}$ , and from the value coming from equation \ref{eq:vartr}  with respect to the integration of the autocorrelation up to its first `zero'. The $\star$ represents the last value at the plateau of $\sigma_\mathcal{V}$ in figures \ref{fig:STDOG} to \ref{fig:STDAG}. b) Evolution of $\mathcal{L}_{Voro}=\eta_C\vert_{\sigma_\mathcal{V}\approx\sigma_{RPP}}$ with $Re_\lambda$. c) Standard deviation of the Vorono\"{i} cells at the last point (filter) on the plateau. In the figures, we are only plotting the AG data for $\delta=0.03$ (see also figure \ref{fig:LINXAG}).}
	\end{figure}

	Furthermore, a comparison between the values estimated from McFadden's equation (taking the last points of the plateaus of $n_s$, and $\sigma_\mathcal{V}$), and the values extracted from $\sigma_\mathcal{V}/\sigma_{RPP}\approx1$ when $\eta_C=\mathcal{L}$ (see figures \ref{fig:LINXAG} and \ref{fig:compx}) points out again that for the AG data, the integral length scale $\mathcal{L}$ could have been originally underestimated by a factor of 2 explaining the discrepancy in $C_\varepsilon$ between the studies of \cite{MoraPRF2019}, and \cite{puga2017normalized}. This underestimation occurs, when integrating the autocorrelation by the method proposed by \cite{puga2017normalized}, as a consequence of the arbitrary choice of the value $\delta$; smaller values converge to a closer value of $\mathcal{L}$, while larger values of $\delta$ reduce the noise while keeping the trends. 
	
	These observations suggest that the integral length scale can also be estimated as $\mathcal{L}=\mathcal{L}_{Voro}=\eta_C\vert_{\sigma_\mathcal{V}\approx\sigma_{RPP}}$. This new definition (see figure \ref{fig:compx2}) seems to be less sensitive to $Re_\lambda$ than the value of $\mathcal{L}$ estimated from equation \ref{eq:vartr}, and to have less dispersion. The values obtained by the latter method show that  for grid turbulence at large $U_\infty$, $\mathcal{L}$ increases with $Re_\lambda$, as previously reported.
	
	On the other hand, and in the context of inertial particles clustering, Monchaux et al. \cite{Monchaux2010} introduced Vorono\"{i} tessellations, and argued that clustering is present when the ratio  $\sigma_\mathcal{V}\vert_\star/\sigma_{RPP}>1$ (where $\sigma_\mathcal{V}\vert_\star$ is the value of $\sigma_\mathcal{V}$ at the plateau observed at small values of $\eta_C$), and its intensity depends on this ratio magnitude. Being the standard deviation mainly set by the `voids' \cite{sumbekova2017preferential} (periods without zero crossings  ), our results show that the degree of clustering of zero crossings increases with $Re_\lambda^{1/3}$ (see figure \ref{fig:stpx}), in agreement with the observations of Mazellier and Vassilicos \cite{mazellier2008turbulence}.
	
	We therefore conclude that while McFadden's equation gives a good estimation of $\mathcal{L}$, a better estimation could be $\mathcal{L}=\eta_C\vert_{\sigma_\mathcal{V}=\sigma_{RPP}}$. The last expression relies on the same hypotheses as McFadden's model, and still has the advantage of producing the value of $\mathcal{L}$ even for non-stationary or non-calibrated data.
	
	\subsubsection{Zero crossing interval PDFs}
	
	The probability density function (PDF) of the inter-arrival distance between zero crossings from turbulent signals, and gaussian processes has been extensively studied in the last decades. Several studies retrieved that this PDF exhibits an exponential behaviour and clustering \cite{sreenivasan1983zero,smith2008fluctuations}. Some studies have reported \cite{chamecki2013persistence,cava2012role} that the onset of such exponential cut-off present in the PDF is due to the randomisation effects (at scales larger than $\mathcal{L}$) that bend the coherent structures present in the flow reducing the probability of larger intervals between zero crossings. 
	
\begin{figure}			
		\begin{center}
			
			\begin{subfigure}[t]{0.48\textwidth}
				\includegraphics[scale=0.3]{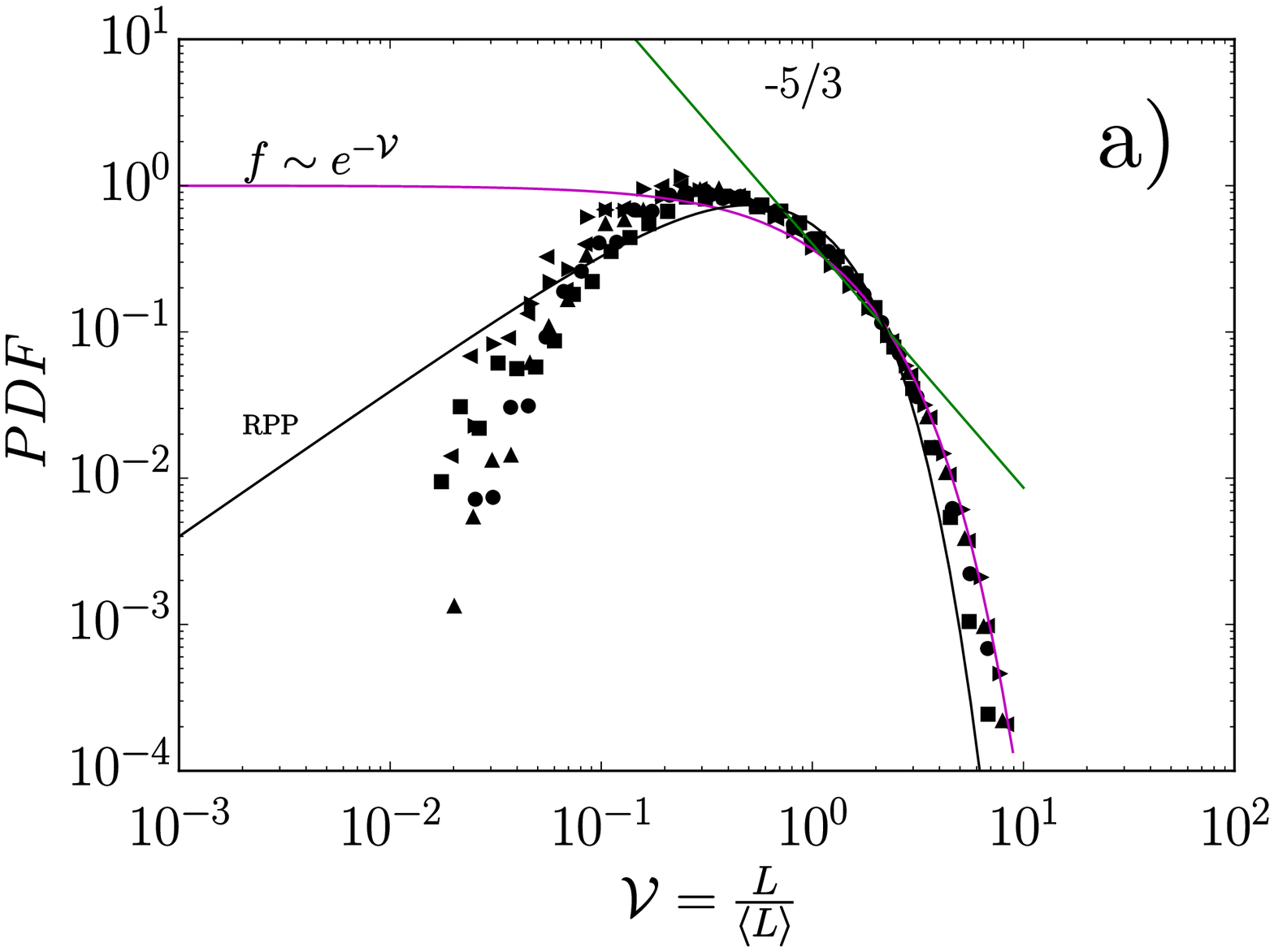}
				\caption{\label{fig:VPFROG}}
			\end{subfigure}
			~
			\begin{subfigure}[t]{0.48\textwidth}
				\includegraphics[scale=0.3]{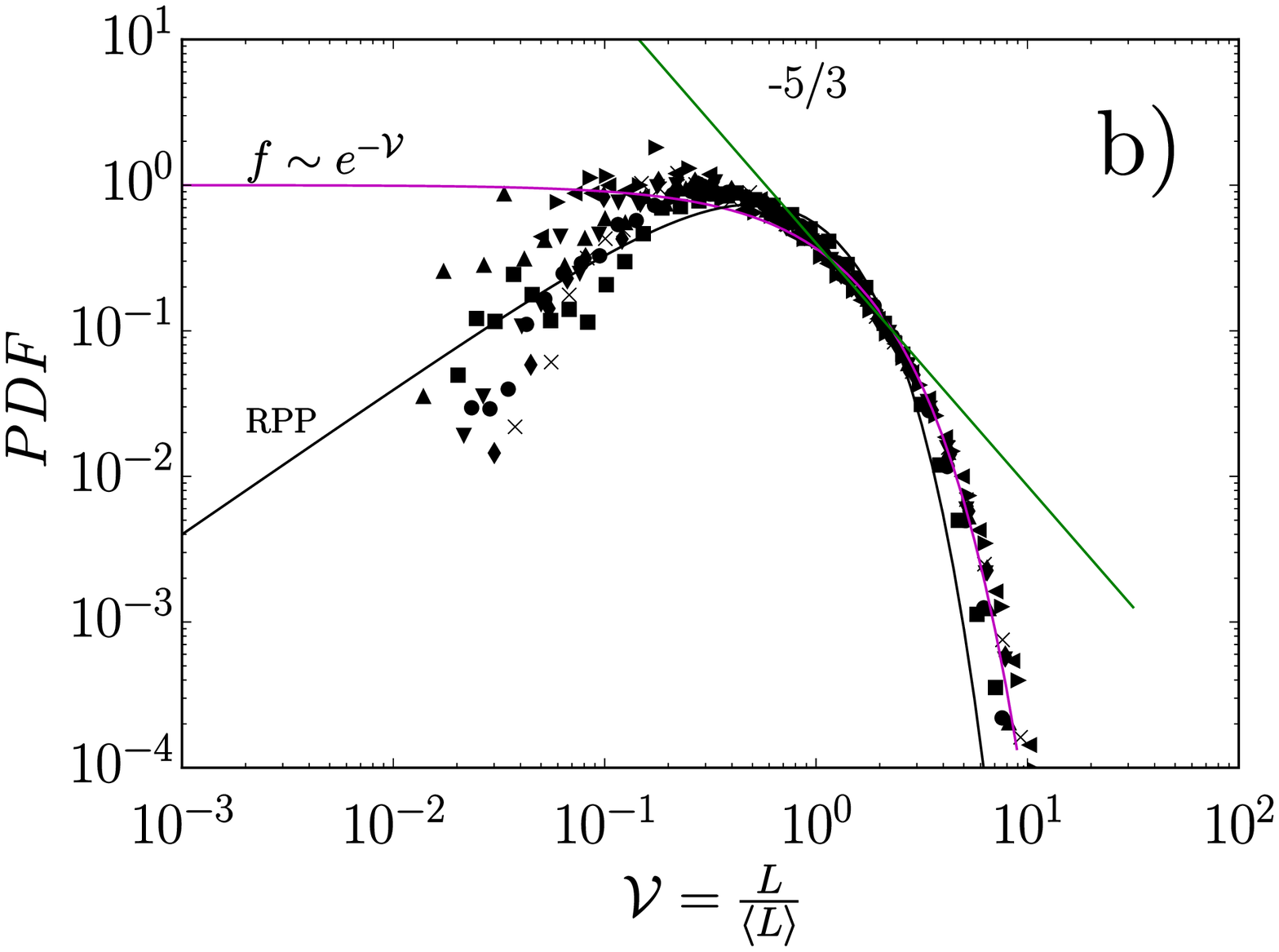}
				\caption{\label{fig:VPFRPG}}
			\end{subfigure}

		\end{center}
				\vspace{-1cm}
		\begin{center}
			\begin{subfigure}[t]{0.48\textwidth}
				\includegraphics[scale=0.3]{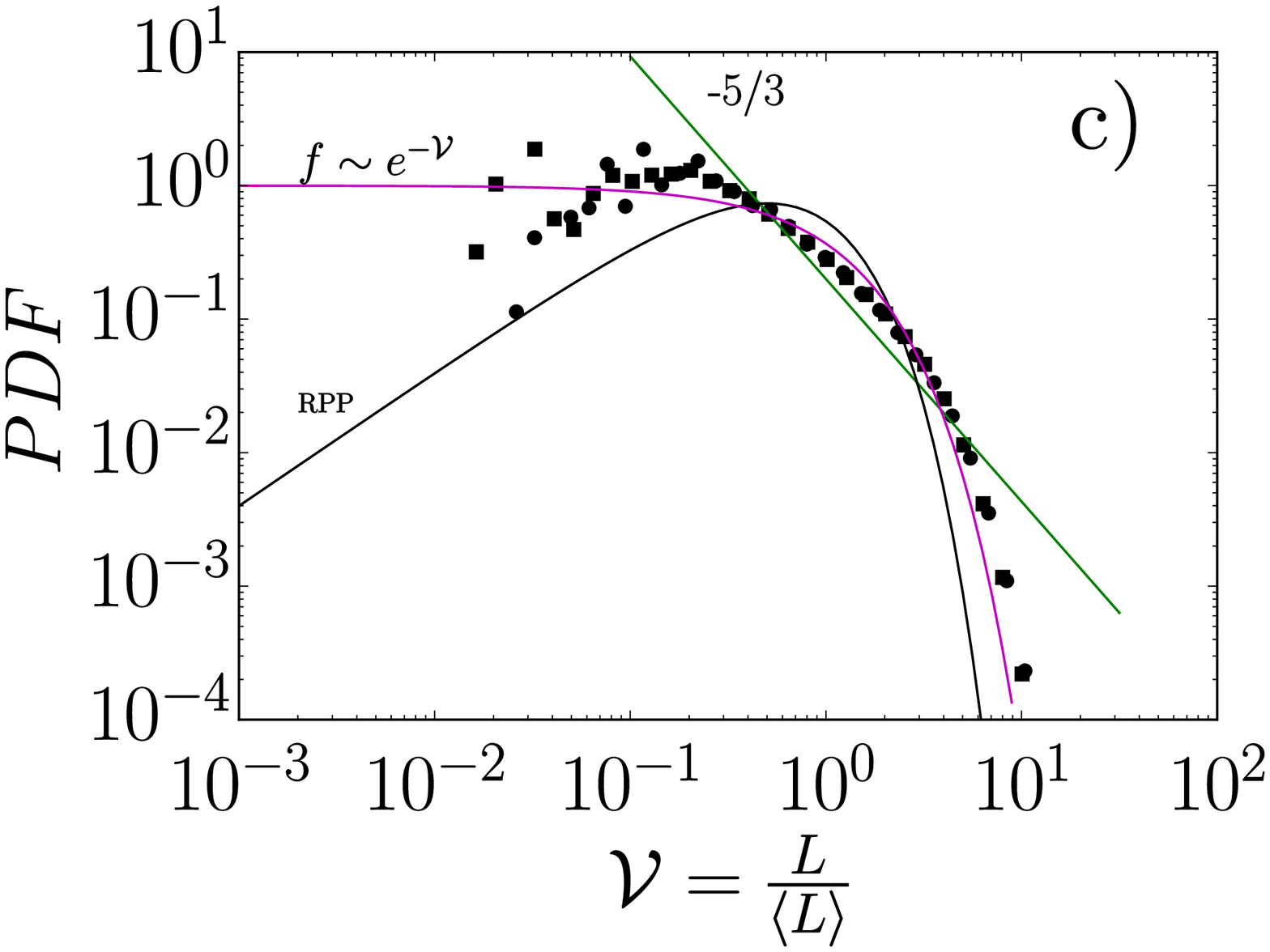}
				\caption{\label{fig:VPFRWK}}
			\end{subfigure}
			~
			\begin{subfigure}[t]{0.48\textwidth}
				\includegraphics[scale=0.3]{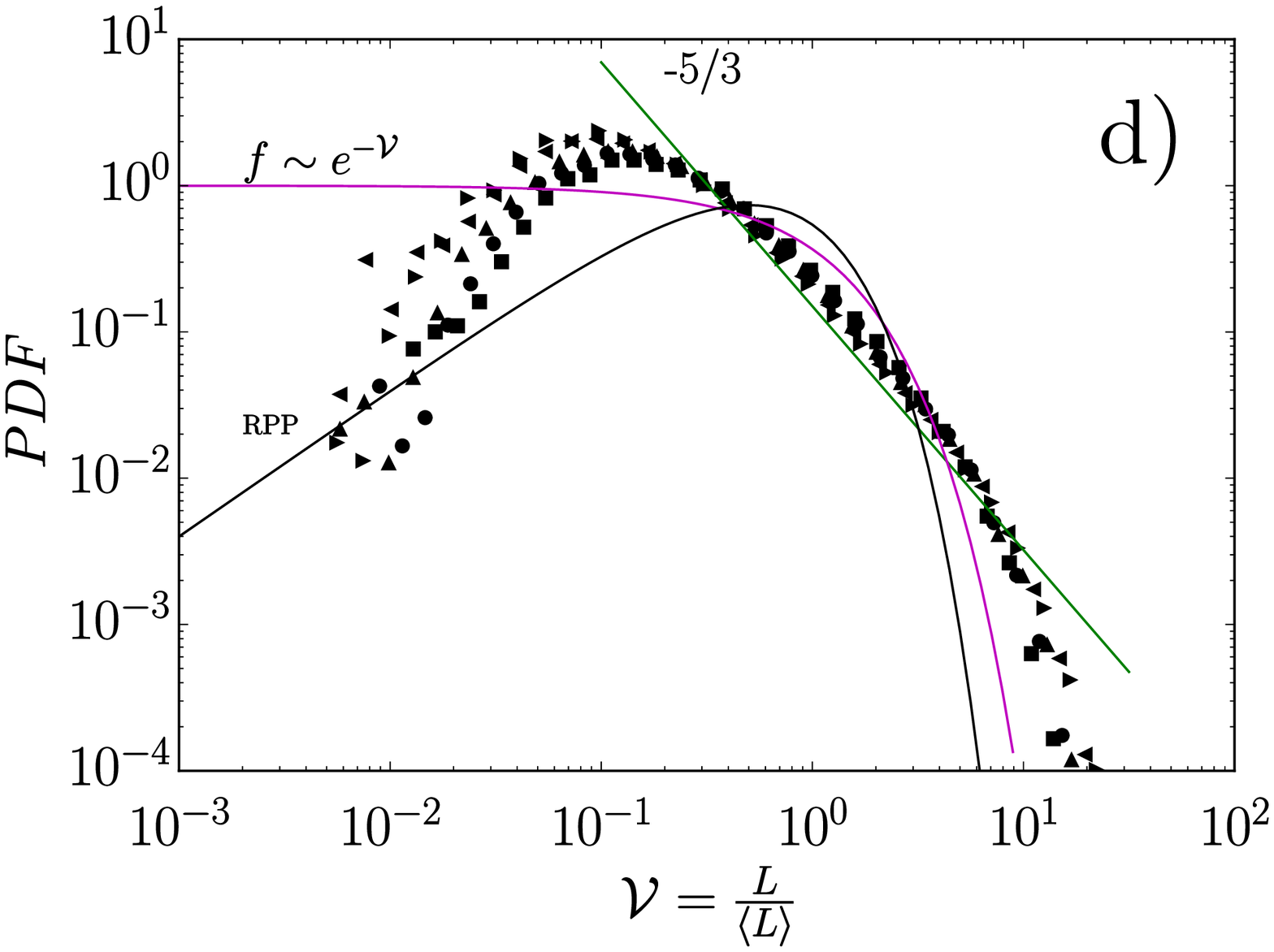}
				\caption{\label{fig:VPFRAG}}
			\end{subfigure}
			
		\end{center}
		\vspace{-0.5cm}
		\caption{PDF of the normalized Vorono\"{i} cell length $\mathcal{V}=L/\langle L\rangle$ for the last data point in the plateau (figures \ref{fig:STDOG} to \ref{fig:STDAG}) and the data found in table \ref{tab:turbParams}. a) OG. b) PG. c) Wakes. d) AG. In the figures the RPP label refers to a random Poisson process with no correlations at any scale \cite{ferenc2007size,Monchaux2010}. }
	\end{figure}

	Our analysis, by means of the 1D Vorono\"{i} tessellation, is consistent with those results: first, we retrieved an exponential cut off transition in our datasets PDFs (see figures \ref{fig:VPFROG} to \ref{fig:VPFRAG}), as well as a power law behaviour (with an exponent close to `-5/3') with increasing $Re_\lambda$. While it is not in the scope of this work, these figures suggest the possibility of using Vorono\"{i} tessellations to do a local analysis of the zero crossings cluster and voids properties (e.g. average cluster size) analogous to those conducted for inertial particles \cite{Monchaux2010}. 

	\section{Concluding remarks}
	
	The velocity autocorrelation function $\rho(r)$ coming from active-grid-generated flows may present a non-decaying behavior that could make ambiguous the estimation of $\mathcal{L}$ by well-established methods. In the previous sections, our analysis of the variance of the distance between zero crossings of the fluctuating velocity via Vorono\"{i} tessellations in conjunction with the theoretical work of McFadden \cite{mcfadden1958axis} allowed us to propose two methods to estimate the integral length scale $\mathcal{L}$. These methods are applicable to hot-wire records coming flows generated by active grids, and thereby, circumvent the problem of the non-standard behavior of $\rho(r)$. They are also consistent with values of $\mathcal{L}$ estimated by traditional methods in several flows: turbulent wakes, and passive grids. The two methods have potential applications in field experiments where calibration could be difficult, or in particle laden flows, where under certain conditions, zero crossing analysis has been used to estimate the energy dissipation rate the presence of inertial particles \cite{MoraPRL2019}.
	Thus, our work shows that all global turbulence parameters (such as $\varepsilon$, $\mathcal{L}$, $\lambda$,...) can be estimated even with a non-calibrated hot-wire, provided that the mean velocity of the flow is known (needed for the Taylor hypothesis).
	
	\section{Acknowledgements}
	
	Our work has been partially supported by the LabEx Tec21 (Investissements d'Avenir - Grant Agreement $\#$ ANR-11-LABX-0030), and by the ANR project ANR-15-IDEX-02. The authors report no conflict of interest.


%
%

\bibliographystyle{spmpsci}      
\bibliography{library.bib}   

%
%

\end{document}